\documentclass{acmsmall}
\usepackage{amssymb}
\makeatletter
\let\@copyrightspace\relax
\makeatother
\usepackage{url}
\usepackage{multirow}
\usepackage{subfigure}

\usepackage[ruled]{algorithm2e}

\SetAlFnt{\small}
\SetAlCapFnt{\small}
\SetAlCapNameFnt{\small}
\SetAlCapHSkip{0pt}
\IncMargin{-\parindent}

\acmYear{2012}
\acmMonth{12}

\begin{document}
\title{Characters and patterns of communities in networks}
\author{
ANGSHENG LI and JIANKOU LI\affil{Chinese Academy of Sciences}
}

\begin{abstract}
A community can be seen as a group of vertices with strong cohesion
among themselves and weak cohesion between each other. Community
structure is one of the most remarkable features of many complex
networks. There are various kinds of algorithms for detecting
communities. However it is widely open for the question: what can we
do with the communities? In this paper, we propose some new notions
to characterize and analyze the communities. The new notions are
general characters of the communities or local structures of
networks. At first, we introduce the notions of {\it internal
dominating set} and {\it external dominating set} of a community. We
show that most communities in real networks have a small internal
dominating set and a small external dominating set, and that the
internal dominating set of a community keeps much of the information
of the community. Secondly, based on the notions of the internal
dominating set and the external dominating set, we define an {\it
internal slope} (ISlope, for short) and an {\it external slope}
(ESlope, for short) to measure the internal heterogeneity and
external heterogeneity of a community respectively. We show that the
internal slope (ISlope) of a community largely determines the
structure of the community, that most communities in real networks
are heterogeneous, meaning that most of the communities have a {\it
core/periphery structure}, and that both ISlopes and ESlopes
(reflecting the structure of communities) of all the communities of
a network approximately follow a {\it normal distribution}.
Therefore typical values of both ISolpes and ESoples of all the
communities of a given network are in a narrow interval, and there
is only a small number of communities having ISlopes or ESlopes out
of the range of typical values of the ISlopes and ESlopes of the
network. Finally, we show that all the communities of the real
networks we studied, have a {\it three degree separation
phenomenon}, that is, the average distance of communities is
approximately $3$, implying a general property of true communities
for many real networks, and that good community finding algorithms
find communities that amplify clustering coefficients of the
networks, for many real networks.
\end{abstract}

\begin{bottomstuff}
The first author is partially supported by
the Hundred-Talent Program of the Chinese Academy of Sciences.
All authors are partially supported
by the Grand Project ``Network Algorithms and Digital Information''
of the Institute of Software, Chinese Academy of Sciences.\\
Author's addresses: Angsheng Li {and} Jiankou Li,
State Key Laboratory of Computer Science, Institute of Software,
Chinese Academy of Sciences, P.O. Box 8718, Beijing, 100190, P.R.China
\end{bottomstuff}
\maketitle
\section{Introduction}
Real networks differ from random graphs in the way that they are
organized with a high level of order. Such an organization results
to remarkable common phenomena of real networks, for instance: the
heavy tail degree distributions, the high clustering coefficients
and the small average distances
etc \cite{barabasi1999emergence,watts1998collective}.
In addition, another remarkable common feature
in various networks is the community structure. Community is an
important notion to disclose the structure of networks, playing the
role in bridging the local vertices and the global network. On one
hand, we could extract communities from a network to study its
internal structure and its relationship with the rest of the network
from the local point of view. On the one hand, we could take each
community as a {\it unit} of the network, to illustrate the
connecting patterns of different communities of real networks
through the distributions of different properties of communities
from the global point of view \cite{de2011exploratory}.

Massive work has been devoted to the study of communities, including
the main definitions of the community problem, algorithms developing
for finding communities, comparison and tests of different
algorithms etc \cite{fortunato2010community}. Leskovec et
al.~\cite{leskovec2009community} analyzed community structures in
large real networks and tried to find the ``best" communities at
various sizes. They showed that the ``best" communities seem to be
characterized by size of $100$.  The distribution of sizes of
communities has also been studied, showing that in some cases, they
have the skewed distribution
\cite{clauset2004finding,newman2004detecting}. The {\it small
community phenomenon} was introduced recently, that is, there are
models, classical or new, such that networks from the models are
rich in small communities, that is, quality communities of small
sizes~\cite{li2011community,li2012smalla}, for which the mechanism
is homophyly.

Intuitively speaking, a community of a network can be interpreted as
a relatively independent and stable unit of the network, and the
rich communities of a network are taken as the {\it local
structures} of the network. This suggests fundamental questions such
as: What can we do with the communities? Are there some characters
of all the communities of a network? What information of the network
can we extract from the communities? What characters of communities
(largely) determine the local patterns of the network? What are the
relationship between the found communities and the true communities?
These questions are widely open in the current state of the art.
This motivates the research in the present paper. For this, we
investigate the following: (1) How to extract central nodes from a
community? (2) How to extract useful information from the
communities? (3) How do communities interact with each other? (4)
How to measure the heterogeneity of a community? (5) What general
properties do the communities (found by a reasonably good algorithm)
have?

By using a variant of the local spectral partitioning algorithm
\cite{andersen2006local}, we find rich communities in real networks.
These networks include collaboration networks, citation networks,
email networks and one benchmark network \footnote{All the data in
this paper can be found from the websites:
\url{http://snap.standford.edu}, or
\url{http://www-personal.umich.edu/~mejn/netdata} and we only
consider the corresponding undirected graphs.}
\cite{girvan2002community,leskovec2007graph}. In collaboration
network a node denotes a scientist and an edge indicates that the
two scientists have coauthored a paper. In the citation networks a
node denotes a paper in some fields and an edge between two papers
indicates that at least one paper has cited the other. Communities
in this networks may correspond to different research groups or
research themes. Two email networks are also used in our study, in
which each node corresponds to an email address and an edge between
nodes $i$ and $j$ represents $i$ sending at least one message to $j$
or $j$ sending at least one message to $i$. A well known benchmark
network of American college football teams complied by Grivan and
Newman~\cite{girvan2002community} is also used. Nodes of the network
represent teams and an edge between two nodes represents that the
corresponding two teams play against each other. The network
contains $12$ true communities, which correspond to $12$ different
conferences that the teams belong to. All networks above have good
community structures so that they are good candidates for
investigating the characters and connecting patterns of local
structures of networks.

We organize the paper as follows. In section \ref{ids_and_eds}, we
propose the notions of {\it internal dominating ratio} and {\it
external dominating ratio} to measure the importance of a subset of
a community. Then we give the definition of internal dominating set
(IDS) and external dominating set (EDS). In section
\ref{local_information_property}, we verify that, the internal
dominating set of a community is much more smaller than the
community and keeps largely the information of the community. In
section \ref{islope_and_eslope}, we define internal slope (ISlope)
and external slope (ESlope) of a community to measure the internal
heterogeneity and the external heterogeneity of the community,
respectively. We analyze the relationship between the structure and
the ISlopes and give the distributions of the ISlopes and the
ESlopes of all the communities of the real networks. In Section
~\ref{other_general_properties}, we analyze more general properties
like average distances, diameters and clustering coefficients of all
the communities for each of the networks. Finally, in section
~\ref{conclusion}, we summarize the conclusions of the paper.

\section{Internal and external dominating sets}
\label{ids_and_eds}

\begin{table}[ht]
\tbl{Statistics of real networks. All the results are calculated
by averaging the corresponding properties of all the communities.
The IDR and EDR are the ratios of centrality of $5$-IDS and $5$-EDS;
the IDN and EDN are the sizes of $0.8$-IDS and $0.8$-EDS
\label{table:new_statistics}}{
\begin{tabular} {|c|c|c|c|c|c|c|}
\hline
Network&IDR&EDR&IDN&EDN&ISlope&ESlope\\
\hline
football&0.99&0.61&2.6&9.3&0.19&0.37\\
\hline
cit\_hepth&0.75&0.49&10&32&0.41&0.54\\
\hline
cit\_hepph&0.73&0.39&12&56&0.5&0.54\\
\hline
col\_astroph&0.93&0.79&3.7&8.1&0.36&0.65\\
\hline
col\_condmat&0.85&0.79&9.6&16&0.42&0.66\\
\hline
col\_grqc&0.94&0.91&3.1&3.9&0.37&0.67\\
\hline
col\_hepth&0.69&0.64&23&27&0.38&0.64\\
\hline
col\_hepph&0.8&0.7&11&16&0.38&0.64\\
\hline
email\_enron&0.93&0.86&3&7.8&0.55&0.68\\
\hline
email\_euall&0.98&0.95&1.7&2.4&0.92&0.89\\
\hline
\end{tabular}
}
\end{table}

Given a community of a network, we may want to extract a small set
of nodes that are more central to the community than the rest of
nodes in the community. Taking the citation network for an example,
we are interested in a small number, $10$ say, of important papers
that are central to the whole community which usually includes
hundreds of papers. In this case, we would hope that with the short
list of key papers, we will not lose any essential information of
the whole community. This analysis of centrality has been studied
for the whole networks, for example, it was shown that a small
fraction of nodes accumulates a large proportion of links in the
networks \cite{newman2004coauthorship}, and that only $20\%$ of
most-linked authors in Economics account for about $60\%$ of all the
links \cite{goyal2006economics}. So there are indeed some nodes
taking the central position in networks. We believe that similar
centrality phenomena occurs in true communities of many real
networks, and that the main goal of community finding algorithms is
to find the true communities of the networks. The question is: what
can we say about the centrality of the communities found by our
algorithms? This would be the first step to understand the
relationship between the true communities and the communities found
by algorithms.

Some centrality measures, initially introduced in social studies,
could be used, for instance, the degree centrality, the closeness
centrality, and the betweenness centrality etc
\cite{freeman1979centrality}. These measures assume a relationship
between the structural position and influential power in group
processes \cite{bavelas1948mathematical}, and are developed and
widely used in the literature~\cite{nicosia2012controlling}. The
mechanism behind this idea is that the centrality of a vertex could
be predicted from its position and the network structure in which it
was embedded as well as from its own characteristics
~\cite{rogers1974sociometric}. Except for these centrality measures,
vertices could also be classified according to their roles within
their communities. Guimer$\grave{a}$ and Amaral decide the role of a
vertex by a {\it within-module degree} $z_i$ and a {\it
participation ratio} $P_i$ and distinguish seven roles that vertices
can play, based on the values of the pair $(z, P)$
\cite{guimera2005functional}.

In this section, we propose the notion of {\it internal and external
dominating sets} of a community by modifying the notion of the
dominating set. The dominating set problem is  classical in graph
algorithms: Given a graph $G=(V, E)$, we say that a set $S\subseteq
V$ is a dominating set if every node $v\in V$ is either an element
of $S$ or adjacent to an element of $S$. The dominating number is
the number of vertices in a smallest dominating set for
G~\cite{allan1978domination,haynes1998funcamentals}.

For a community, we distinguish two roles that nodes can play in a
community, as an internal role and an external role, measured by
links within and outside of the community respectively. For a subset
of a given community, its {\it internal dominating ratio} (IDR, for
short) is defined as follows.

Let $C$ be a community, $S$ be a subset of $C$, $N(S,C)$ be the
neighbors of $S$ within community $C$. Then we define the {\it
internal dominating ratio of $S$ in $C$}, written by IDR,  as
follows:
\begin{equation}
{\rm IDR}(S) = \frac{|S\cup N(S,C)|}{|C\cup N(C, C)|} = \frac{|S\cup
N(S,C)|}{|C|} \label{eq:idr}
\end{equation}

The dominating ratio has been used previously to measure the social
centrality in social networks~\cite{freeman1979centrality}. Our
internal dominating ratio (IDR) measures the importance of a group
of nodes in a community, and thus it can be seen as a general format
of degree centrality of communities.

 Following the definition above, we consider two
problems: 1) when given a number $k$ (usually small), we want to
find a subset $S$ of size $k$ with $\max\{{\rm IDR}\}$, in which
case, we call this subset a $k$-IDS; 2) when given a real number $p$
in $[0,1]$, we want to find a subset $S$ whose IDR is bigger than
$p$ with the minimum number of nodes, in which case, we call this
subset a $p$-IDS.

Similarly to IDR, we give the definition of {\it external dominating ratio}
(EDR). Let $C$ be a community, $S$ be a subset
of $C$, $N(S, \bar C)$ be
the neighbors of node $v$ that are outside of $C$. Then the external
dominating ratio ( EDR) of $S$ in $C$ is defined as follows:
\begin{equation}
{\rm EDR}(S) = \frac{|N(S, \bar C)|}
              {|N(C, \bar C)|}
\end{equation}

\begin{figure}
\centering
\includegraphics[width=3in]{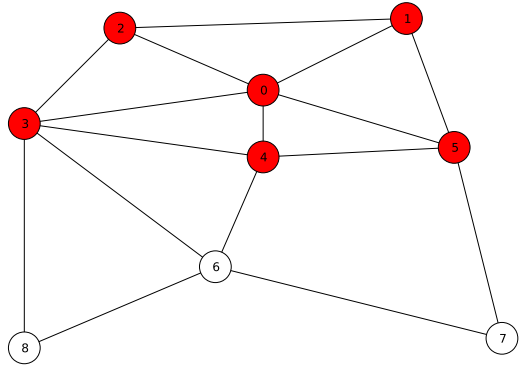}
\caption{A community to illustrate the IDR and EDR. The red nodes
are from one community. Let $S = \{v_3, v_4 \}$ be a subset of
this community,
$N(S, C) = \{v_0, v_2, v_5\}$, $N(S, \bar C) = \{v_6, v_8\}$, 
$N(C, \bar C) = \{v_6, v_7, v_8\}$, so IDR(S) = (2 + 3) / 6 = 5/6, EDR(S) = 2/3}
\label{fig:idc_edc}
\end{figure}

We also give the notations $k$-EDS and $p$-EDS similarly. Figure
\ref{fig:idc_edc} is an example of the IDRs and EDRs. From the
definitions, we notice that we are not using the notion of
classic dominating set~\cite{allan1978domination,haynes1998funcamentals},
instead, we introduce two parameters $k$
and $p$ to define the general format of dominating sets. We
emphasize that the classification are based on nodes positions in a
community. By definition, it is conceivable that nodes in the IDS
are more important for the function and stability of the community,
and that nodes in the EDS mainly take charge of the communication
between the community and the nodes outside of the community.

The dominating problem is an NP-complete decision
problem~\cite{haynes1998funcamentals}. Here we introduce a simple
greedy algorithm to find the $p$-IDS and $p$-EDS, where $G$ is a
graph, $C$ is a community and $p$ is a real number in $ [0, 1]$.

\begin{algorithm}[t]
\SetAlgoNoLine
\KwIn{Graph $G$, community $C$, and a real number $p\in [0, 1]$.}
\KwOut{The $p$-internal dominating set $S$.}
Let $G_C$ be the induced subgraph
of vertices $C$ from $G$. Set $S = \emptyset$\;
\Repeat{${\rm IDR}(S) \leq p$}
       {
 Let $v$ be a node in $C\setminus S$ such that $v$ has the
 maximal number of neighbors in $C\setminus (S\cup N(S))$, where
 $N(S)$ is the neighbors of $S$ in $G_C$\;
 $S\leftarrow S\cup \{v\}$ \;
       }
\caption{Finding $p$-\{{\rm IDS}\}}
\label{alg:finding_p_ids}
\end{algorithm}

\begin{algorithm}[t]
\SetAlgoNoLine
\KwIn{Graph $G$, community $C$, and a real number $p\in [0, 1]$.}
\KwOut{The $p$-external dominating set $S$.}
Set $S = \emptyset, N(S, \bar C) = \emptyset$, where
$N(S, \bar C)$ is the number of neighbors of $S$ outside of $C$\;
\Repeat{${\rm EDR}(S) \leq p$}
       {
 Let $v$ be a node in $C\setminus S$ such that $v$ has the
 maximal number of neighbors in $N(C, \bar C)\setminus N(S, \bar C)$\;
 $S\leftarrow S\cup \{v\}$\;
 $N(S,\bar C)\leftarrow N(S, \bar C)\cup N(v, \bar C)$\;
       }
\caption{Finding $p$-\{{\rm EDS}\}}
\label{alg:finding_p_eds}
\end{algorithm}

Given a number $p$ between $0$ and $1$, we could find the $p$-\{{\rm
IDS}\} and $p$-\{{\rm EDS}\} by using the above algorithms.
Similarly when given a small number $k$, we could calculate the
$k$-\{{\rm IDS}\} and $k$-\{{\rm EDS}\} by using the same algorithm
with slight modification of the terminating condition. In our
experiment, we set $k = 5$ when calculating the $k$-\{{\rm IDS}\}
and the $k$-\{{\rm EDS}\}, and set $p = 0.8$ when calculating the
$p$-\{{\rm IDS}\} and $p$-\{{\rm EDS}\}, see Table
\ref{table:new_statistics} for details.

From Table \ref{table:new_statistics}, we observe that only five
nodes could dominate most of the members of the communities from
both internal and external sides, that the internal dominating
ratios of $5$ internally central nodes are larger than the external
dominating ratio of $5$ externally central nodes, for each of the
networks, that external connecting patterns of the communities are
more decentralizing than that of the internal connecting patterns,
for each of the networks, that it only needs at most $10$ nodes to
internally dominate at least $80\%$ of the whole community, that it
needs at most $32$ nodes to externally dominate $80\%$ of the
outgoing links of the communities, and that external dominating
numbers are larger than the internal dominating numbers for all
communities and for all the networks.

 In summary, we have that most communities have a small internal
 dominating set, and a small external dominating set, which is
 slightly larger than the internal dominating set of the
 corresponding communities, on the average, for all the networks.

\section{Extracting Local Information}
\label{local_information_property}

In the last section, we verify that most communities have a small
internal dominating set, and a small external dominating set. The
questions are: How much information of a community is preserved in
the dominating set of the community? How to extract essential
information of a community from the small dominating sets?

In this section, we verify that the internal dominating sets (IDSs)
indeed preserve essential information of the communities. We verify
this result by predicting and confirming keywords of papers in a
citation network.

 We say that a paper has keywords, if its
authors have explicitly list its keywords, and does not have
keywords, otherwise.

Keywords of papers play an important role in information retrieval.
In many citation networks, there is a huge number of papers whose
keywords are not listed by their authors, which is an obstacle for
people to sufficiently use the networks\footnote{We implement the
verification for just one citation network, because this is the only
available network in which titles, abstract of papers, and keywords
of a small number of papers are included. Most networks have a
topological structure with nodes and edges only.}.

In the citation-hepth networks, there are about $27,770$ papers, in
which only $10\%$ or so have keywords. Predicting and confirming the
missing keywords for the other papers are obviously significant for
information retrieval.

Given a community $C$ in a citation network, we predict and confirm
keywords for papers in $C$ by the following procedure.

\begin{algorithm}[t]
\SetAlgoNoLine
\KwIn{Graph $G$, community $C$, and keyword dictionary $Dic$
}
\KwOut{Papers with predicting keywords}
Calculate $p$-IDS or $k$-IDS of $C$\;
Suppose that $L = \{k_1, k_2, \cdots, k_i\}$ are listed keywords
from the IDS with descending order according to their popularity in
$C$. For a given paper $P$ in $C$ whose keywords are not listed in
the network, for each $j\leq i$, if $k_j$ appears in either the
title or the abstract of paper $P$, we say that $k_j$ is a predicted
and confirmed keyword of $P$\;
\caption{Predicting keywords using internal dominating set}
\label{alg:one}
\end{algorithm}

We choose parameter $p=0.8$, run the algorithm on the citation
network, and report the results in Table
\ref{table:predict_keyword}. The first column of
Table~\ref{table:predict_keyword} presents the number of keywords we
used for the prediction and confirmation for each communities, that
is, the length $i$ of $L$ in the algorithm, the second column of the
table are numbers of papers whose keywords have been predicted and
confirmed corresponding to different lengths of $L$ in the first
column.

From Table \ref{table:predict_keyword}, taking the first row of the
table for example, we know that if we use the most popular $5$
keywords appearing in the IDS of each of the communities, then there
are $13,283$ papers in the network whose keywords are predicted and
confirmed. As the number of keywords used in the algorithm, i.e.,
the lengths of $L$ in the algorithm, becomes larger, we can predict
and confirm keywords for more papers, that is up to $14,691$ papers.
The results show that the IDS is much smaller than the corresponding
community and that the IDS preserves much information of the
corresponding community. From the experiment, it is conceivable that
in practical applications, it is sound to recommend the IDS of a
community instead of the whole community which is usually much
larger. The result above is unexpectedly good. We believe that this
property may hold for many other networks other than citation
networks, that is, the internal dominating set of a community keep
essential information of the community. More importantly, the
essential information of the internal dominating set of a community
can be easily extracted.

\begin{table}
\tbl{Using 0.8-IDS to predict keywords in citation network hepth
\label{table:predict_keyword}} 
{
\begin{tabular} {|c|c|}
\hline
Keyword Number&Predicted Paper Number\\
\hline
5&13283\\
\hline
10&13906\\
\hline
15&14375\\
\hline
20&14592\\
\hline
25&14641\\
\hline
30&14647\\
\hline
35&14654\\
\hline
40&14691\\
\hline
45&14691\\
\hline
50&14691\\
\hline
\end{tabular}
}
\end{table}

\section{Internal and External Slopes }
\label{islope_and_eslope} In the last section, we show that most
communities have a small IDS and a small EDS, and that the small IDS
of a community preserves much information of the community.

In this section, we will show that the IDS and EDS of a community
usually take the central positions in the community with low degree
nodes around them, so that the community forms a core/periphery
structure.

Intuitively speaking, if all nodes in a community have equal
position, i.e., the regular graph or a random graph, then they are
homogeneous; if nodes in a community form a core/periphery
structure, i.e., the star-like graphs, then they are heterogeneous.
Our main question is: How do the IDS and EDS of a community
reflect the homogeneity or the heterogeneity of the community?

Before answering this question, we look at the power law
distribution. It was shown that most networks follow a power law
distribution~\cite{barabasi1999emergence}, meaning that the
number of nodes of degree $k$ is proportional to $k^{-\beta}$.
A power law distribution of power exponent $\beta$, 
which is typically lying in the range $2 < \beta < 3$,
measures the heterogeneity of a network.
However it is nontrivial to estimate the exponent $\beta$,
especially for small networks, and not all networks follow the power
law distribution \cite{clauset2009power}. Most communities are
small, although they may have heavy tail degree distributions, it is
not clear whether they have power law distributions. More seriously,
even if the communities have power law distributions, fluctuations
caused by the small sizes of communities may make the result
inaccurate, and the number of communities is large, it is hard to
characterize the power law distributions of all the communities.
Therefore the power exponent $\beta$ is not suitable to measuring
the heterogeneity of all the communities of a network. Another
measure is to notice the relationship between the number of
dominating set and the degree distribution. In fact, it was shown
that the more heterogeneous  the degree distribution of a network
is, the smaller the number of dominating set
is~\cite{jose2012dominating}. This suggests that the internal and
external dominating sets are closely related to the heterogeneity of
the communities.

\begin{figure}
\centering
\subfigure[ISlope = 0]
{\label{fig_islope_a}
\includegraphics[width=2.5in]{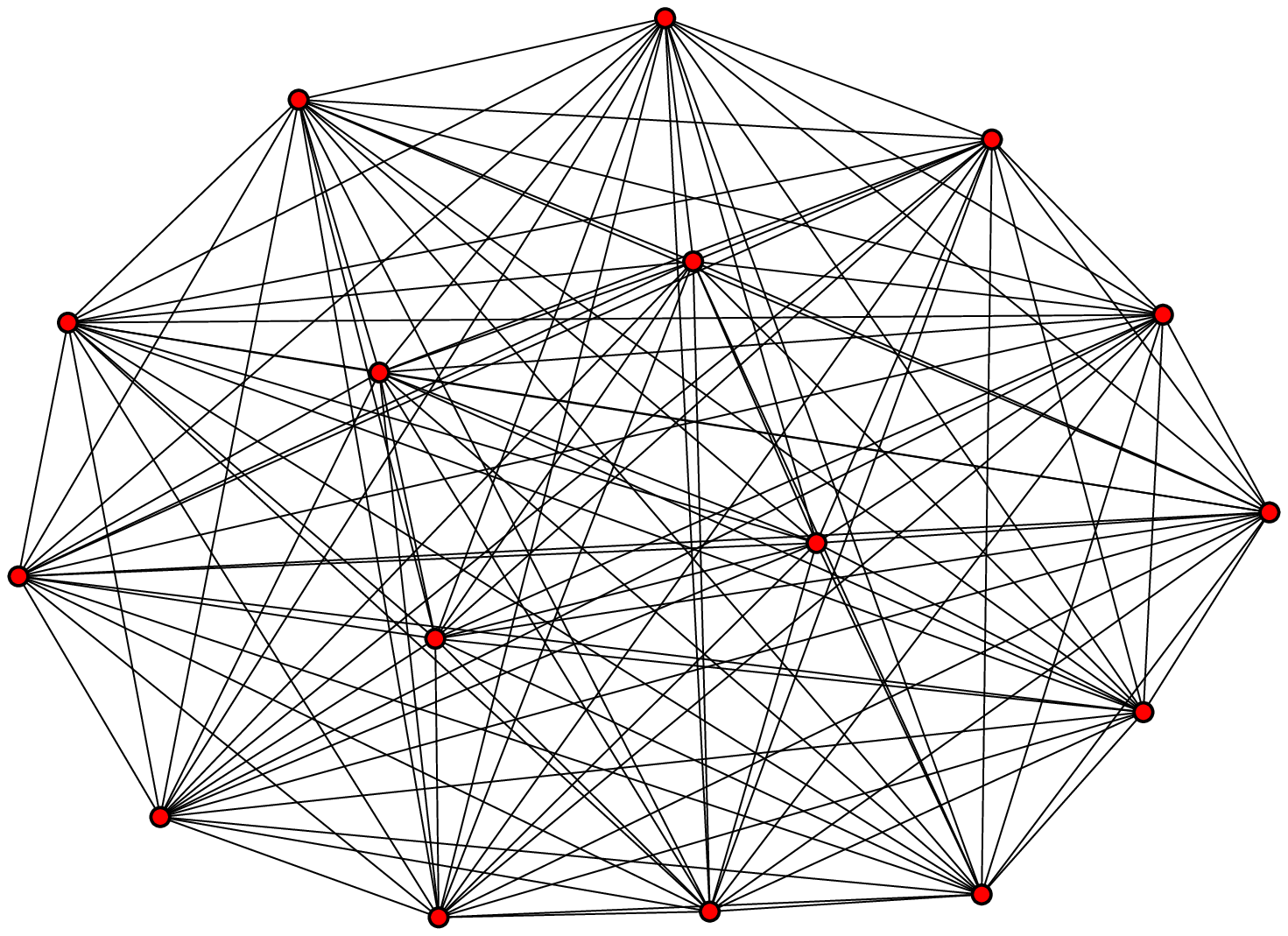}
}
\subfigure[ISlope = 0.41]
{\label{fig_islope_b}
\includegraphics[width=2.5in]{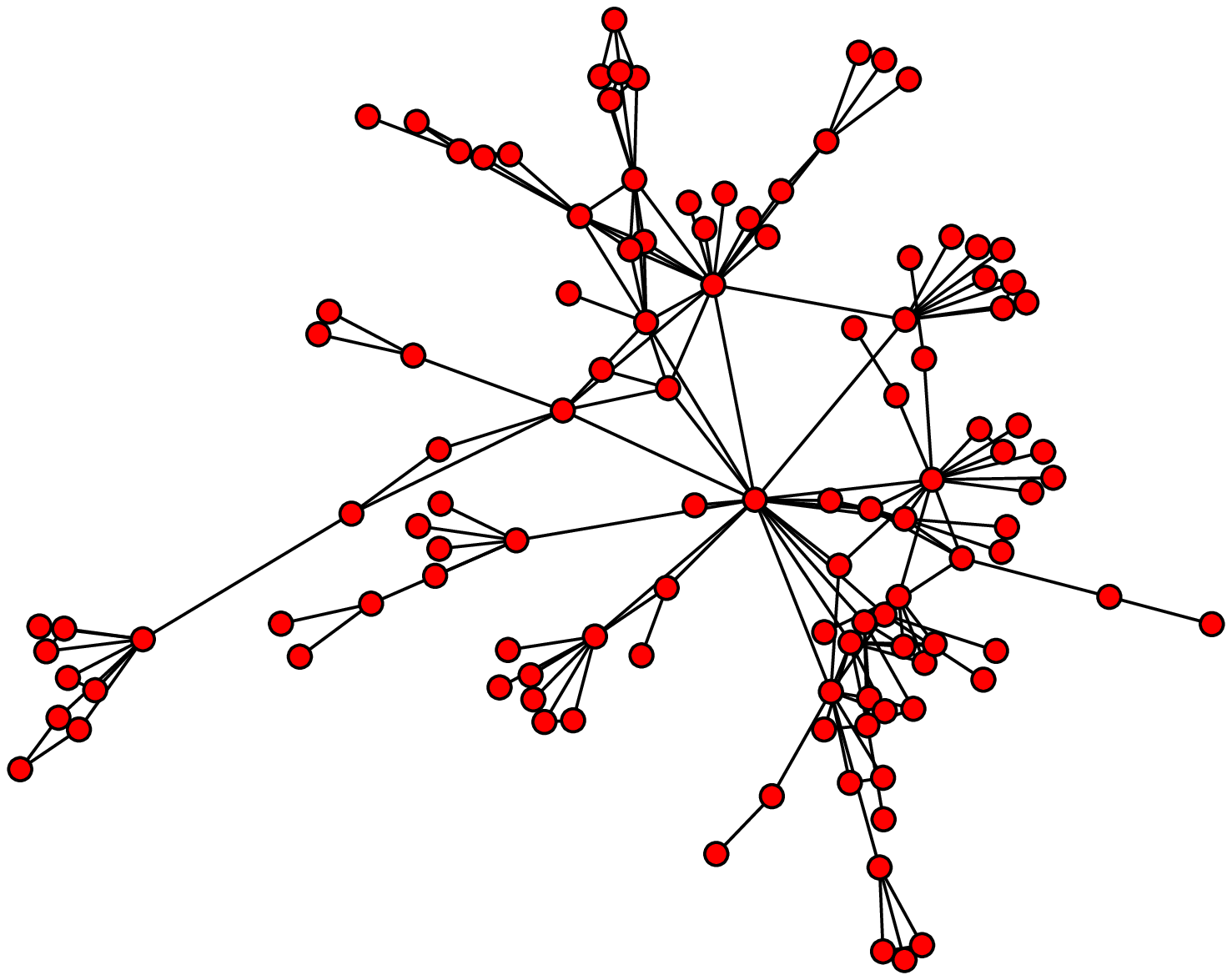}
}
\subfigure[ISlope = 0.72]
{\label{fig_islope_c}
\includegraphics[width=2.5in]{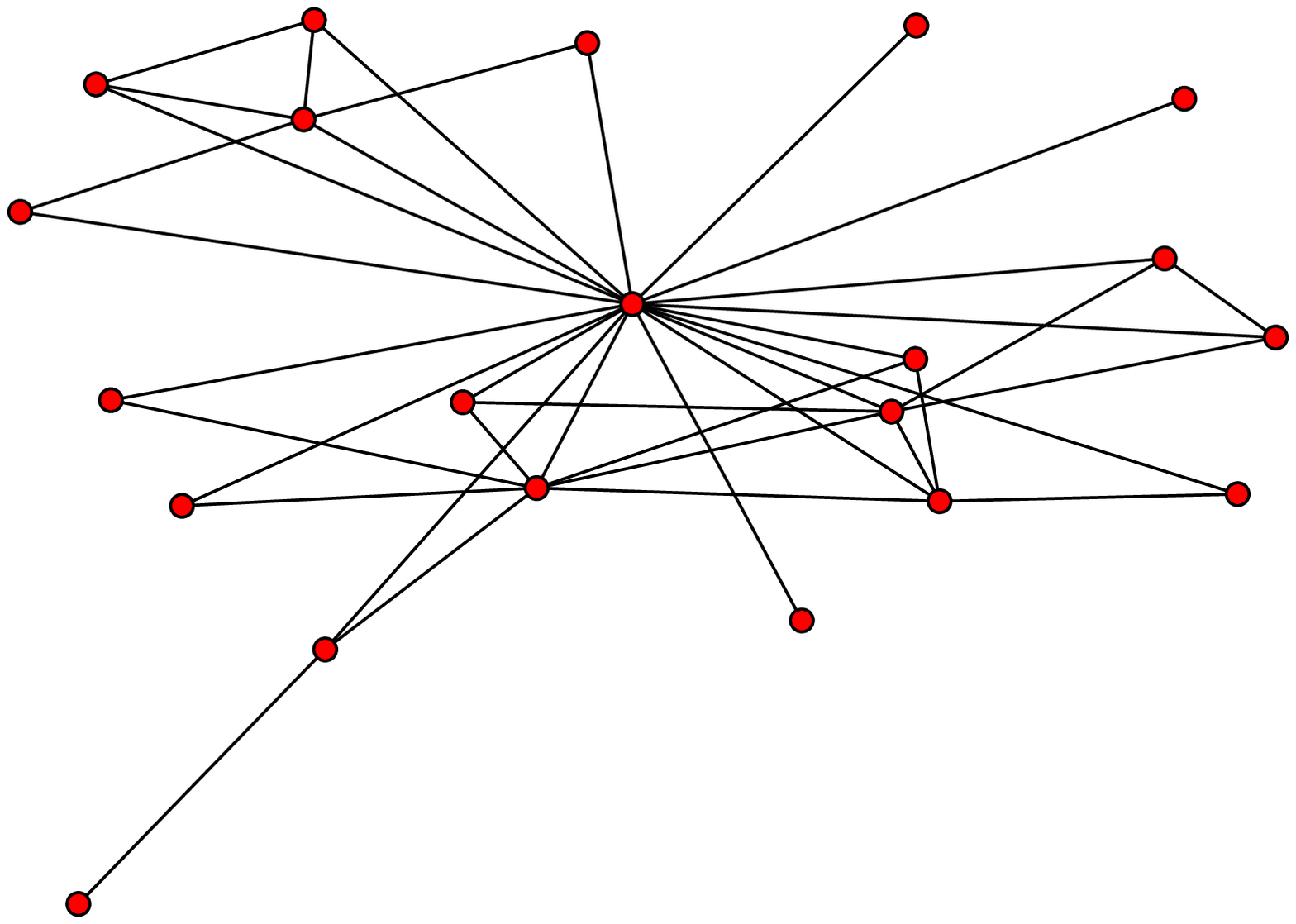}
}
\subfigure[ISlope very near 1]
{\label{fig_islope_d}
\includegraphics[width=2.5in]{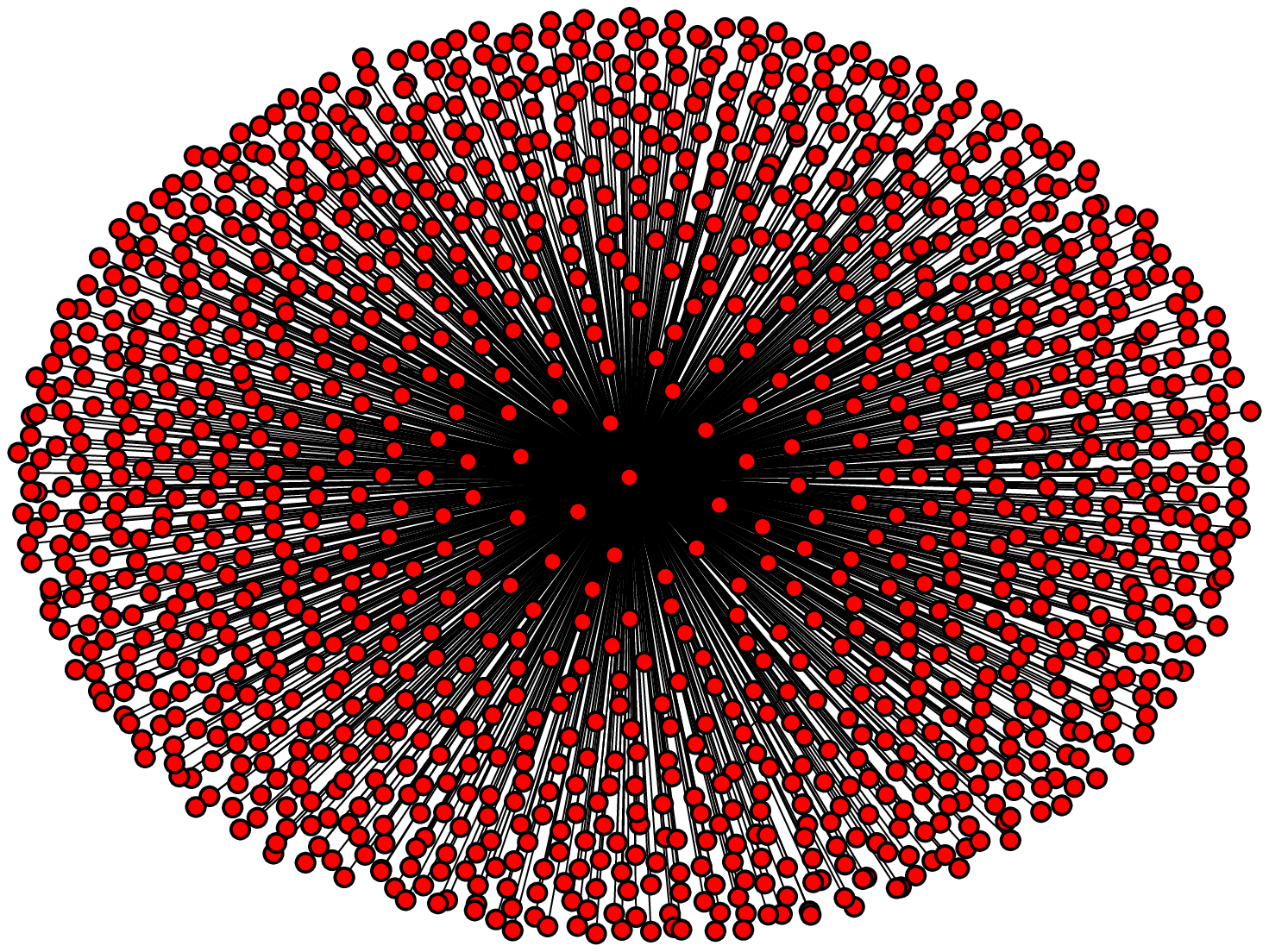}
}
\caption{Real communities to illustrate ISlope.
All of them except (d) are from collaboration grqc network
and (d) is from email\_enron network. In each figure, red nodes
come from the same community.}
\label{real_fig_islope}
\end{figure}

We now measure the heterogeneity of communities by the internal and
external dominating sets of communities. 
See figure \ref{fig_islope_a} in which case the community is
homogeneous. All members of the community have equal
position, and any single node could dominate the whole community.
From the dominating number, we could not know the heterogeneity of
the community. So the dominating set itself is insufficient to measure
the homogeneity and heterogeneity of a community. To solve this
problem, we use the internal dominating ratio (IDR) of the internal
dominating set (IDS), together with the expectation internal
dominating ratio (IDR) of random selection of nodes of the same
size as that of the IDS.

We define the {\it internal slope} (ISlope, for short) and {\it
external slope} (ESlope, for short) of a community to measure the
internal and external heterogeneity (or the core/periphery
structure) of the community. Intuitively, the ISlope of a community
is to measure the distance
between the community and regular graphs or star-like graphs, and
the ESlope of a community is to illustrate whether the community is
connected with the rest of the community evenly or through a small
number of nodes like a funnel.

Let $C$ be a community, $p\in [0, 1]$ be a real number. Suppose that
$K$ is the size of the $p$-IDS of $C$, that $S$ be the $p$-IDS of
$C$, and that $\mathcal{V} = \{V_1, V_2, \cdots, V_M\}$ is the set
of all subsets of $C$ of size $K$. Then define the {\it internal
slope of $C$}, written by ${\rm ISlope} (C)$ as follows:

\begin{equation}
{\rm ISlope}(C) = {\rm IDR}(S) - \frac{\sum_{X\in \mathcal{V}}{\rm
IDR}(X)}
                        {|\mathcal{V}|}
\end{equation}

The ISlope of a community represents the difference between the {\it
internal dominating ratio} of the most central nodes and the
expectation internal dominating ratio of random choices of nodes of
the same size. It measures the homogeneity and heterogeneity
(core/periphery structure) of the community from the internal point
of view. We extract some communities of real networks found by our
algorithm in Figure \ref{real_fig_islope}. From these figures
we can observe that the ISlopes and ESlopes of the communities
largely reflect the homogeneity and the heterogeneity of the
corresponding communities.

\begin{figure}
\centering
\subfigure[ESlope = 0]
{\label{fig_eslope_a}
\includegraphics[width=2.5in]{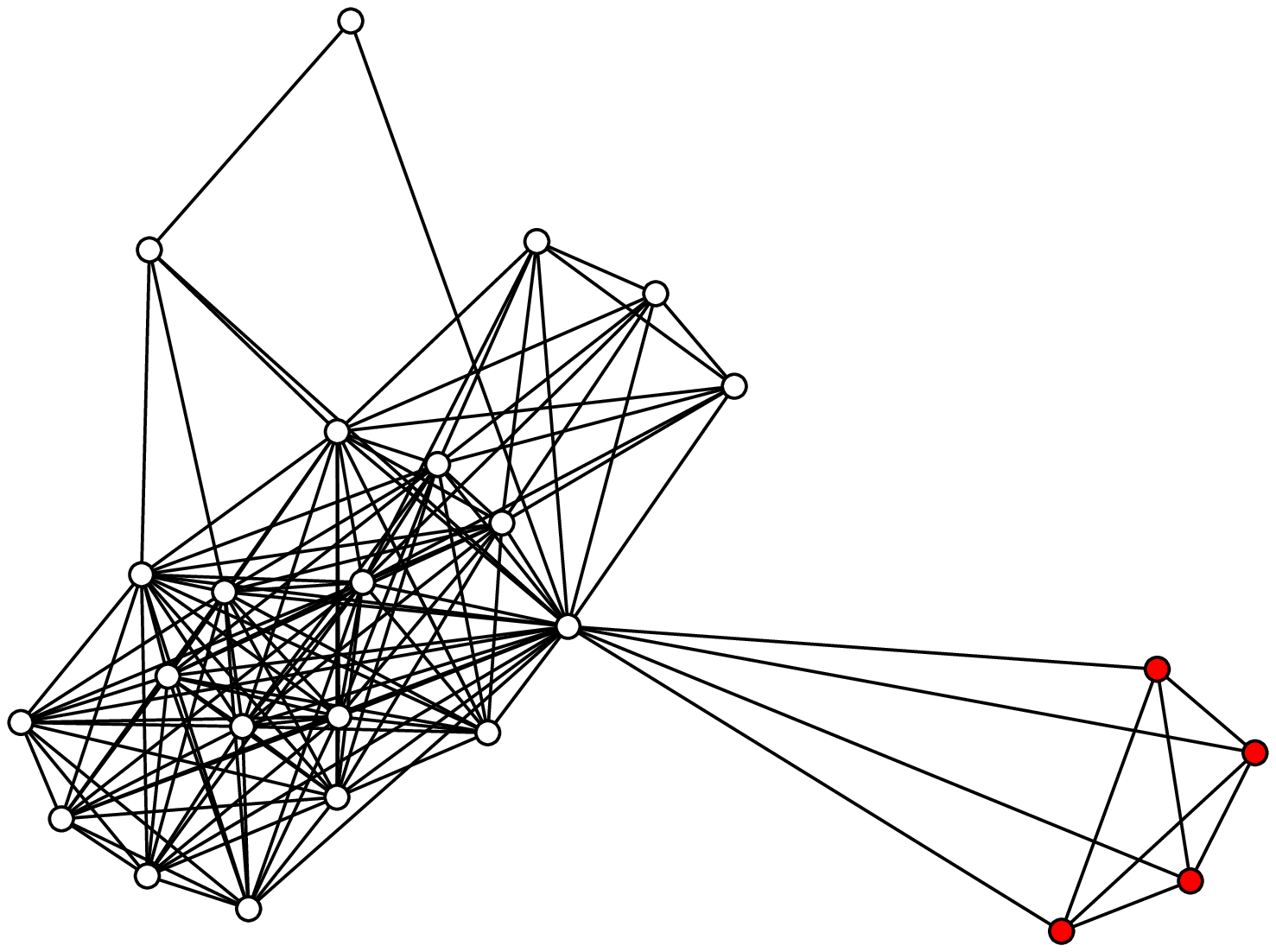}
}
\subfigure[ESlope = 0.31]
{\label{fig_eslope_b}
\includegraphics[width=2.5in]{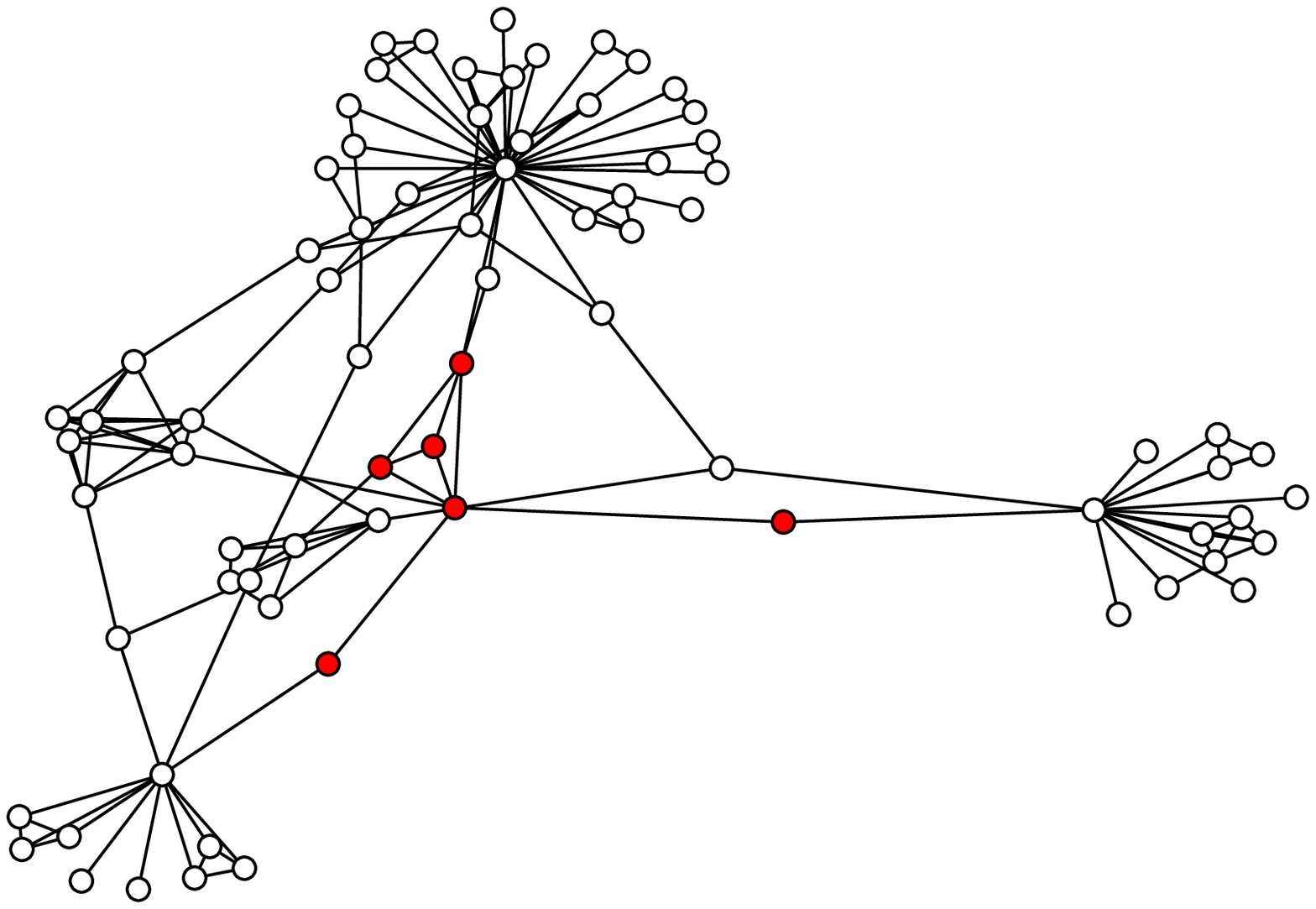}
}
\subfigure[ESlope = 0.52]
{\label{fig_eslope_c}
\includegraphics[width=2.5in]{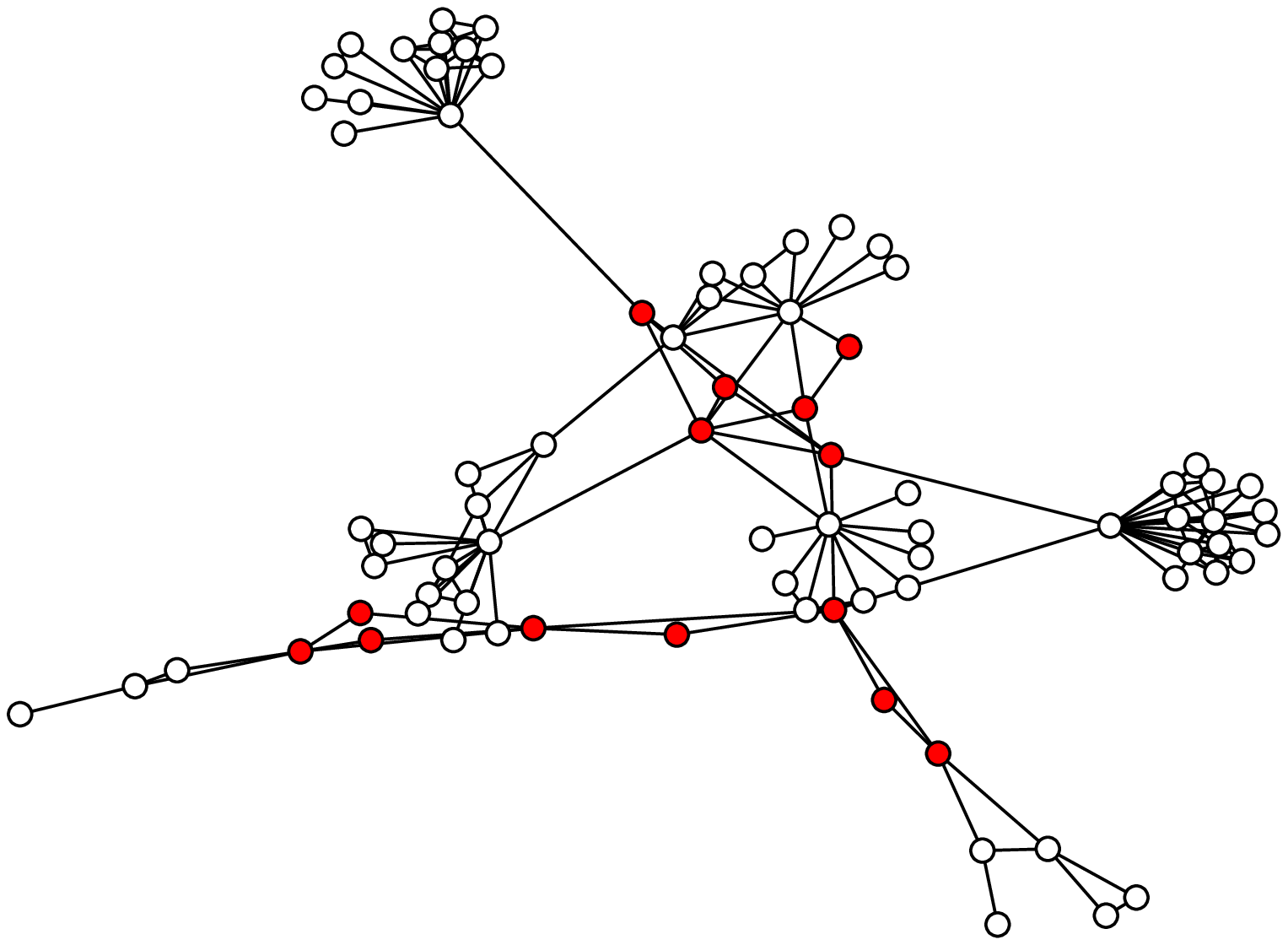}
}
\subfigure[ESlope = 0.97]
{\label{fig_eslope_d}
\includegraphics[width=2.5in]{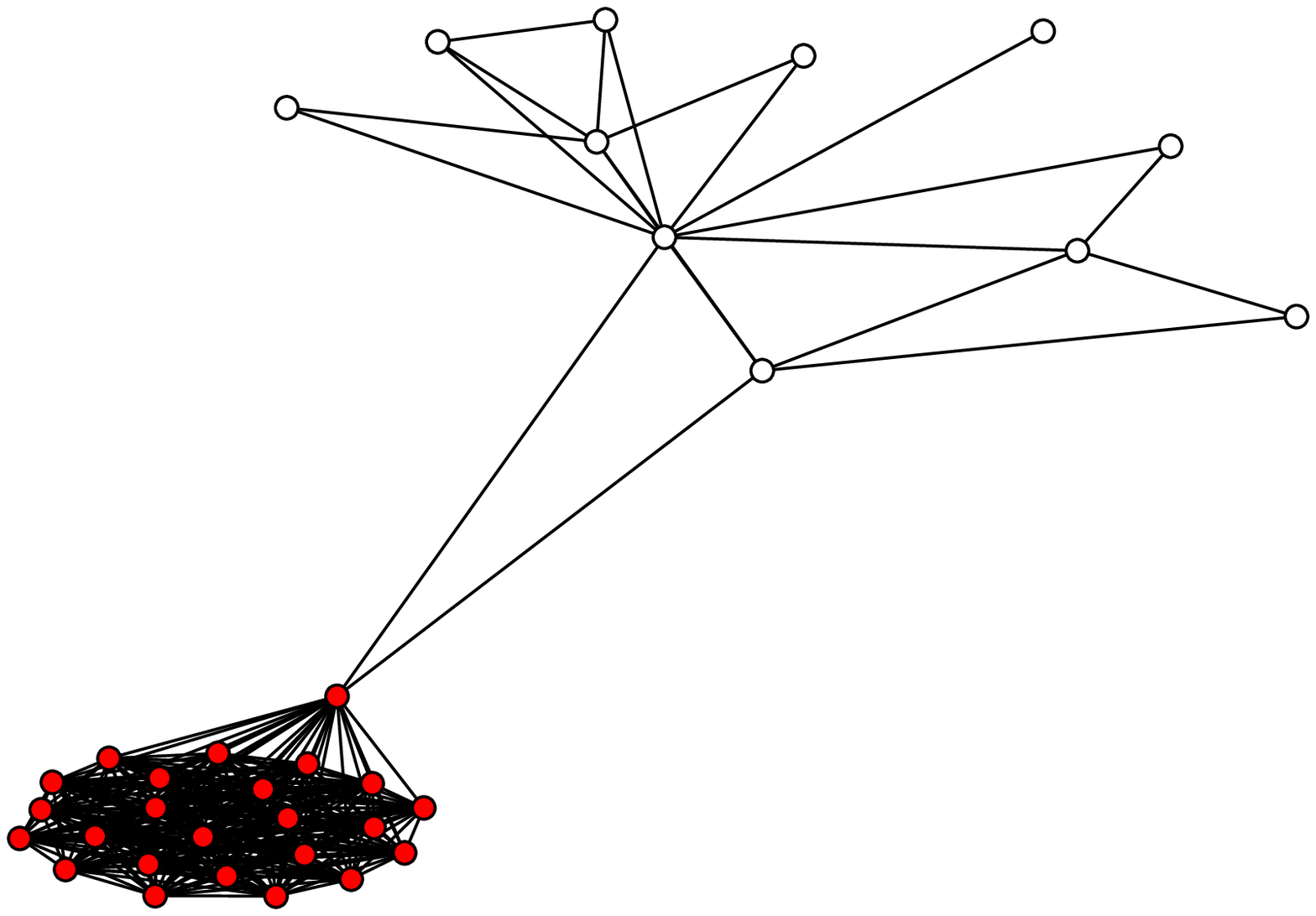}
}
\caption{Real communities to illustrate ESlope.
All of them are from collaboration grqc network.
In each figure, red nodes come from the same community.}
\label{fig_eslope}
\end{figure}

By observing Figure~ \ref{real_fig_islope}, we know that the structures
of communities are closely related to the corresponding ISlopes of
the communities. In particular, in Figure \ref{fig_islope_a}, all
nodes have equal position and a single node could dominate the whole
community; in Figure \ref{fig_islope_b}, there are some central
nodes with periphery nodes around; in Figure \ref{fig_islope_c}, the
central position of one node is more obvious, and the structure is
a star-like graph; in Figure \ref{fig_islope_d}, the community is a
star graph with a hub in the center, and the ISlope of the community
is very near $1$. Notice that a star graph is the most heterogeneous
community, in which the hub in its center is the most important
node. In summary, we observe that the smaller the ISlope of a
community is, the more homogeneous a community is, and that on the
contrary, the larger the ISlope of a community is, the more
heterogeneous a community is, and that the ISlope of a community
roughly reflects the pattern or structure of the community.

Similarly to ISlope, we define the {\it external slope of  a
community} (ESlope) to measure the external heterogeneity of the
community. By using the ESlope of a community, we are able to
examine the pattern that nodes in a community connect nodes outside
of the community. Whether or not nodes in a community connect the
rest of the community through a small number of representatives or
evenly through most members.

It has been shown that in a collaboration network, most people in
the network (theme, or topic) contact people in the network through
just one or two of their best-connected
collaborators~\cite{newman2004coauthorship,newman2001bscientific}.

Our results show that such a funneling pattern of connections from a
community to outside of the community is very popular in all the
communities of a network, for a wide range of real networks.

Let $C$ be a community, $p\in [0, 1]$ be a real number. Suppose that
$K$ is the size of a $p$-EDS of $C$, that $\mathcal{V} = \{V_1, V_2,
\cdots, V_m\}$ is the set of all subsets of $C$ of size $K$. Then we
define the {\it external slope of C} (${\rm ESlope} (C)$) as
follows:

\begin{equation}
{\rm ESlope}(C) = {\rm EDR}({\rm EDS}) - \frac{\sum_{Y\in
\mathcal{V}}{\rm EDR}(Y)}
                        {|\mathcal{V}|}
\end{equation}

The ESlope of a community represents the difference between the
external dominating ratio of the most central nodes and the
expectation {\it external dominating ratio} of random selection of
nodes of the same size.

Figure \ref{fig_eslope} illustrates different connecting patterns of
communities with different ESlopes. In these figures, we also keep
the neighbors and the neighbors of neighbors of the community to
highlight their connecting patterns. In figure \ref{fig_eslope_a},
all members have equal position to connect with nodes outside of the
community. Some nodes only have internal links, while others have
both external and internal links in figure \ref{fig_eslope_b}. Also,
some nodes play the role of bridge in linking nodes in and outside
of its community in figure \ref{fig_eslope_c}. At last, figure
\ref{fig_eslope_d} shows a community in which only one node is the
bridge. All other members communicate with the outside world through
this node. The ESlope indeed identifies different connecting
patterns of how communities connect with each other.

Table \ref{table:new_statistics} gives the average ISlopes and
ESlopes of all the communities of various networks. Except for the
football and the email\_euall, all other networks have similar
ISlopes and ESlopes with ESlopes larger than ISlopes, on the
average. ISlope and ESlope of a community quantify the
core/periphery structure of the community. Our results indicate that
such structures are universal in real networks and that real
networks tend to avoid communities of either regular or star-like
graphs and have structures with ISlopes and ESlopes in some fixed
interval, that is, the ISlopes are roughly in $[0.35, 0.55]$ and the
ESlopes in $ [0.5, 0.7]$.

These results pose a question that why networks tend to have such
structures. We try to explain these as follows: For a community, it
is possible that some key nodes are essential to its formation and
evolution. On one hand, it is unusual to have a community with all
members having equal position for a long period of time. On the
other hand, the key nodes of a community should not be too strong or
too weak since otherwise, the community structure may be fragile. It
is intuitive that if the central nodes of a community breakdown,
then the community structure would not exist any more. Therefore too
big ISlopes or ESlopes and too small ISlopes or ESlopes will both go
ill with the  evolution of communities. The structures of typical
communities of a real network may be a compromise between the
effectiveness and robustness of the communities. We conjecture that
the ESlopes may largely determine the evolution of communities,
which needs to be further investigated ( in our on going project).

\begin{figure}
\centering
\subfigure[ISlope\_football]
{\label{fig_islope_distri_football}
\includegraphics[width=2.5in]{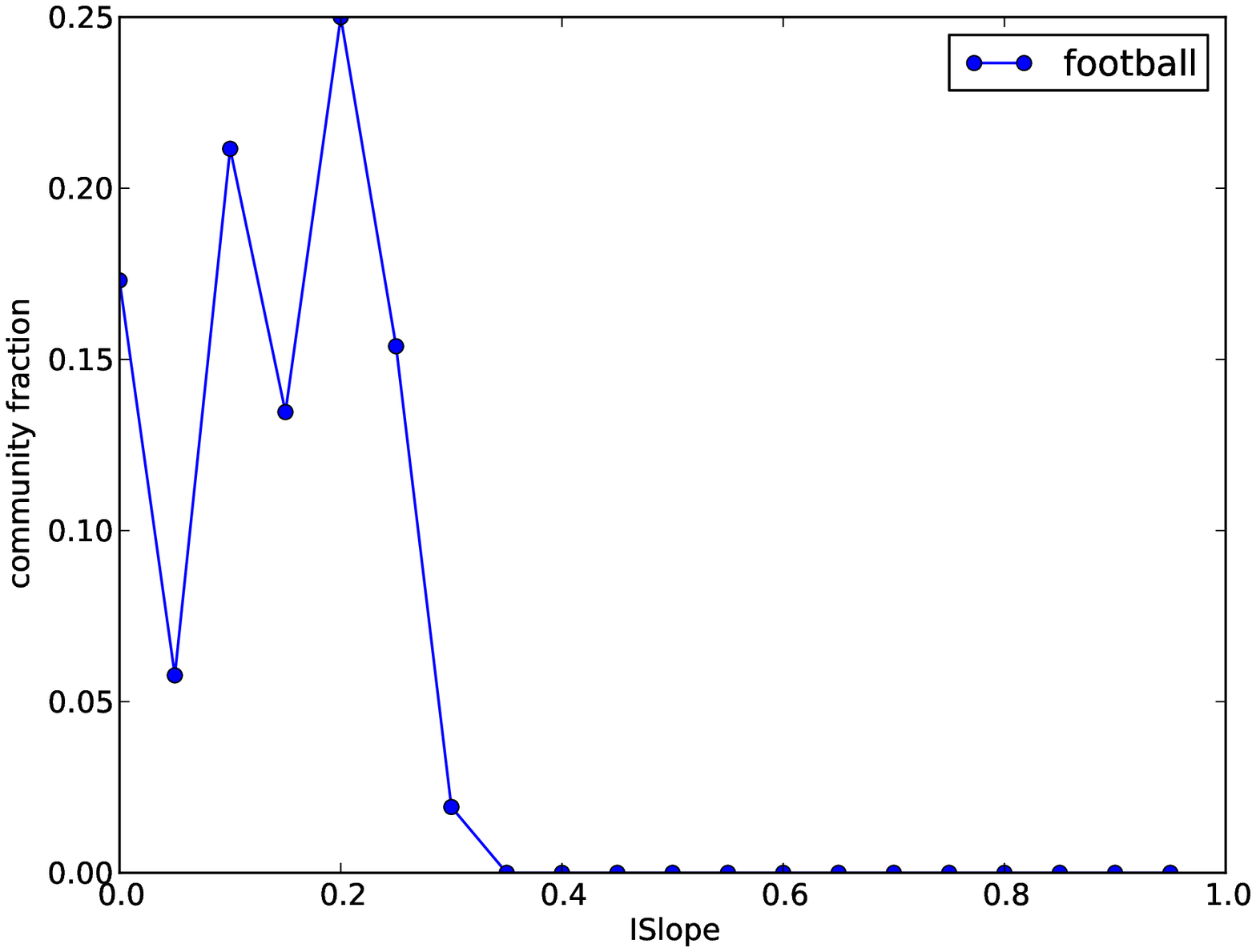}
}
\subfigure[ISlope\_citation]
{\label{fig_islope_dirstri_citation}
\includegraphics[width=2.5in]{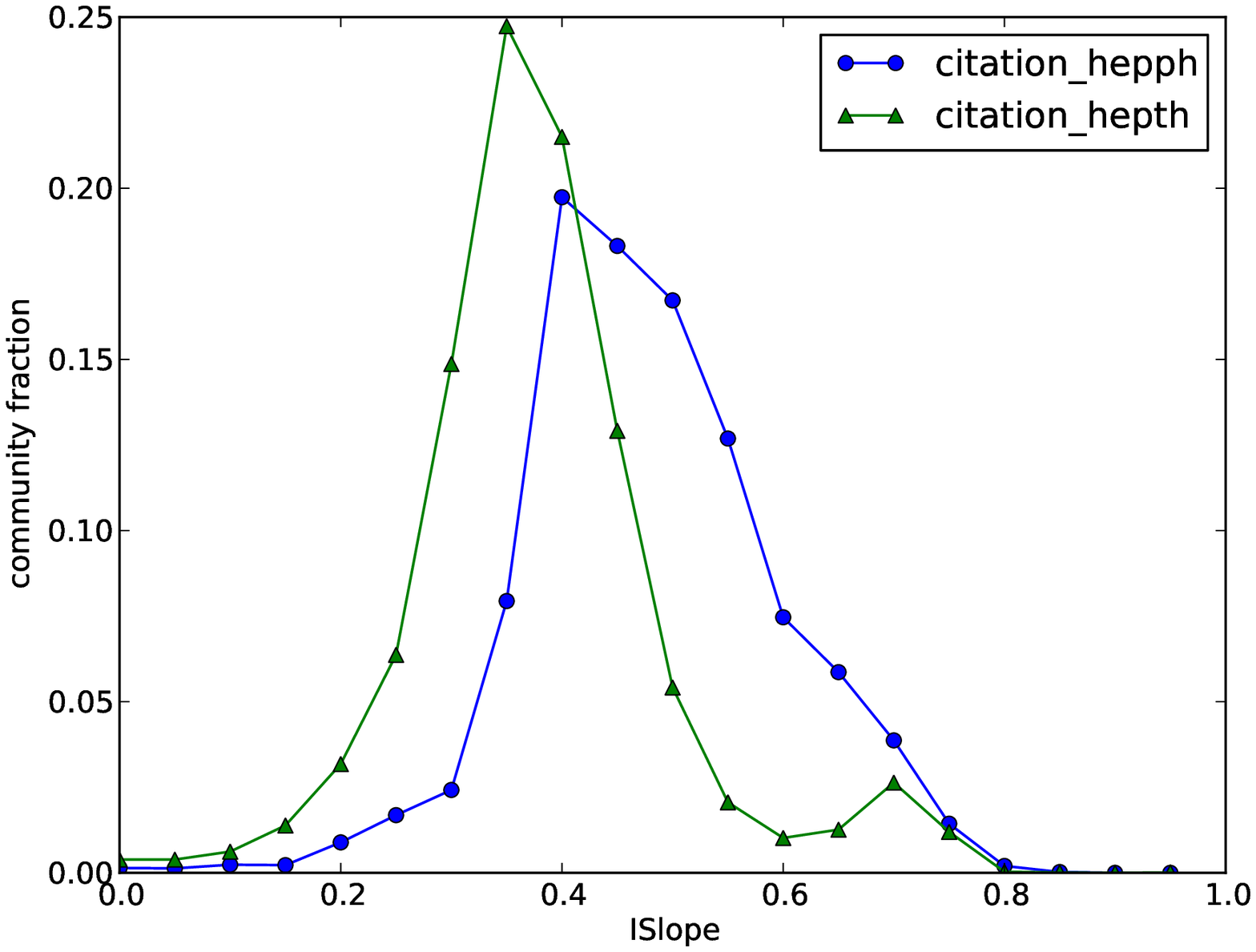}
}
\subfigure[ISlope\_collaboration]
{\label{fig_islope_distri_collaboration}
\includegraphics[width=2.5in]{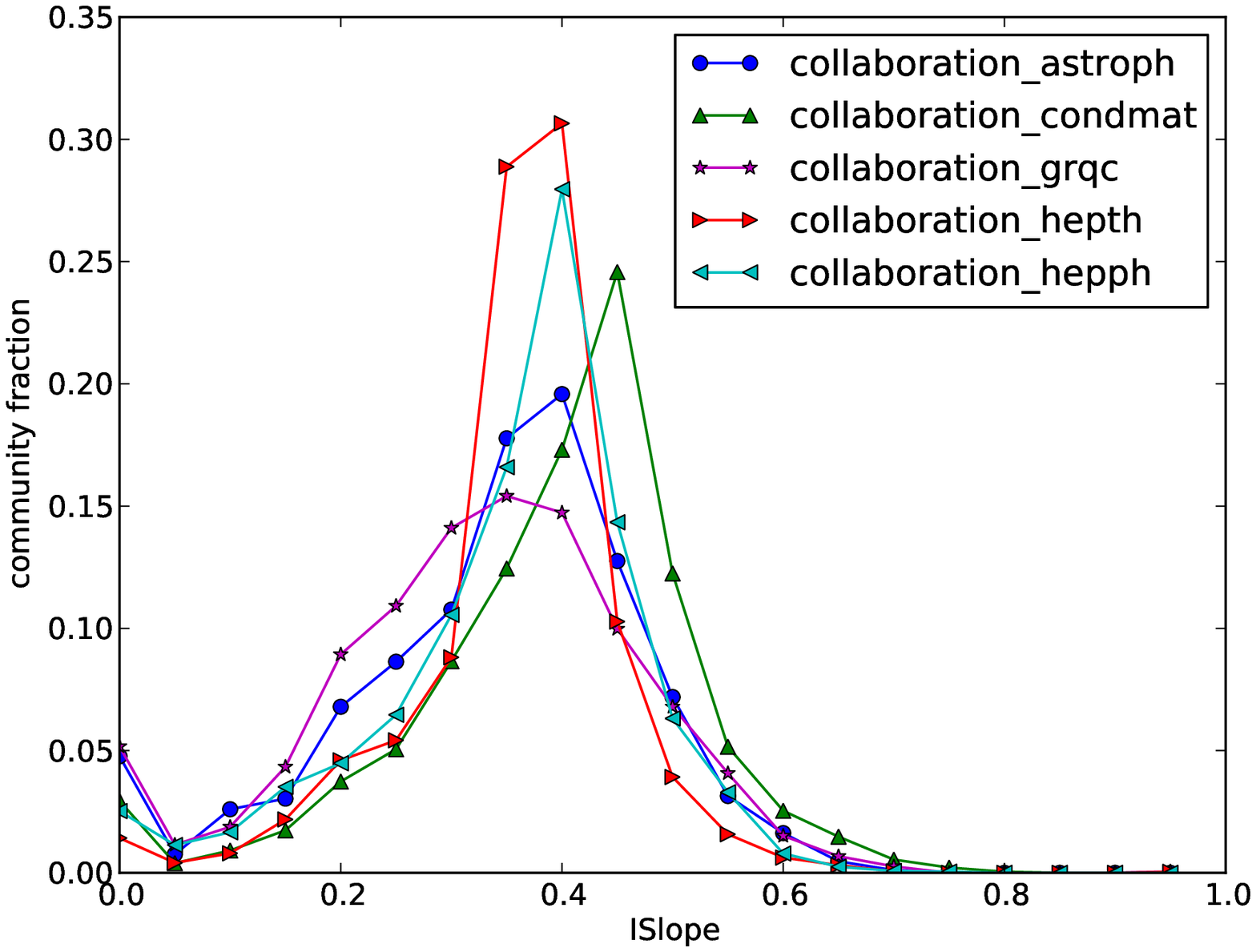}
}
\subfigure[ISlope\_email]
{\label{fig_islope_distri_email}
\includegraphics[width=2.5in]{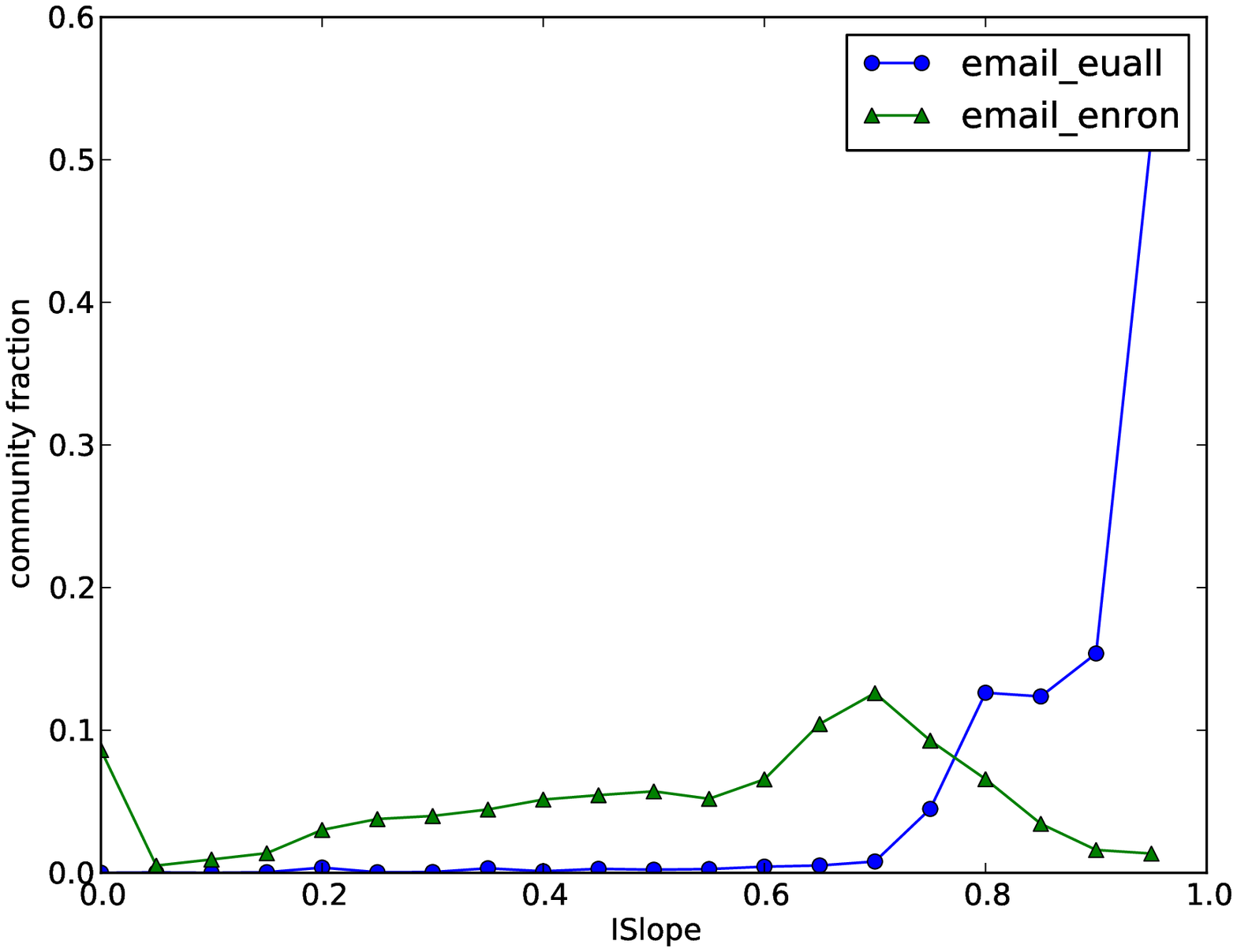}
}
\caption{Distribution of communities' ISlope}
\label{fig_islope_distri}
\end{figure}

\begin{figure}
\centering
\subfigure[ESlope\_football]
{\label{fig_eslope_football}
\includegraphics[width=2.5in]{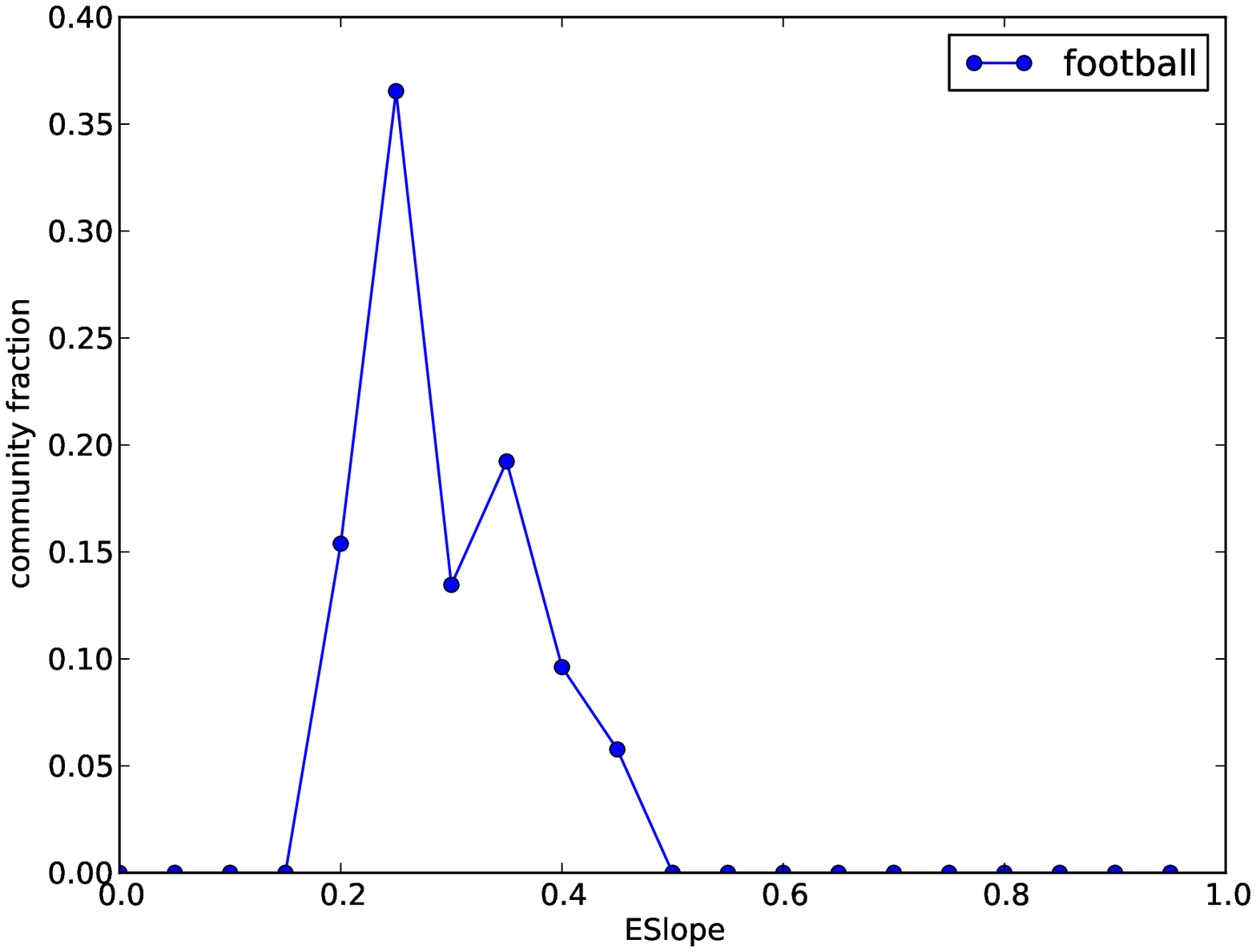}
}
\subfigure[ESlope\_citation]
{\label{fig_eslope_citation}
\includegraphics[width=2.5in]{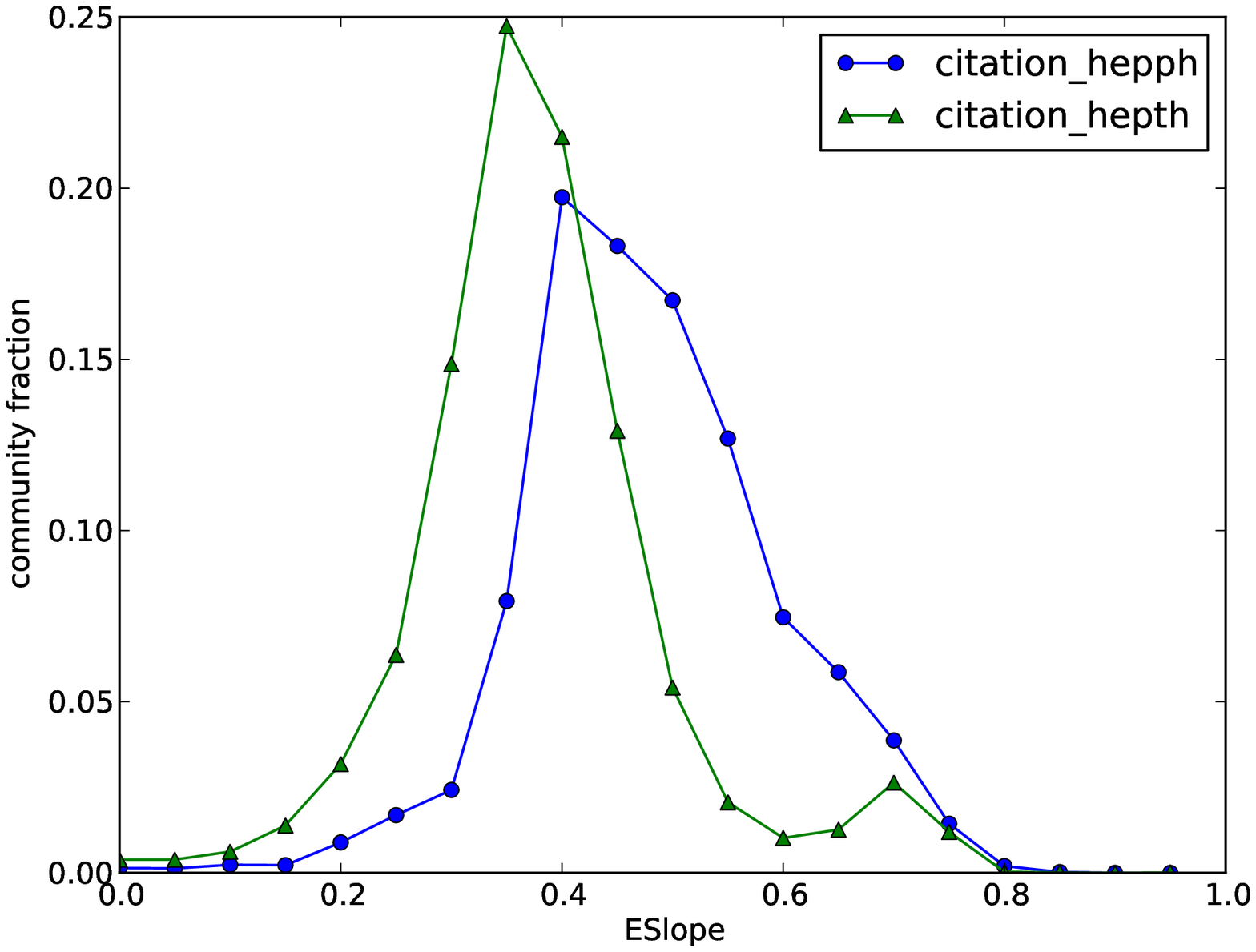}
}
\subfigure[ESlope\_collaboration]
{\label{fig_eslope_collaboration}
\includegraphics[width=2.5in]{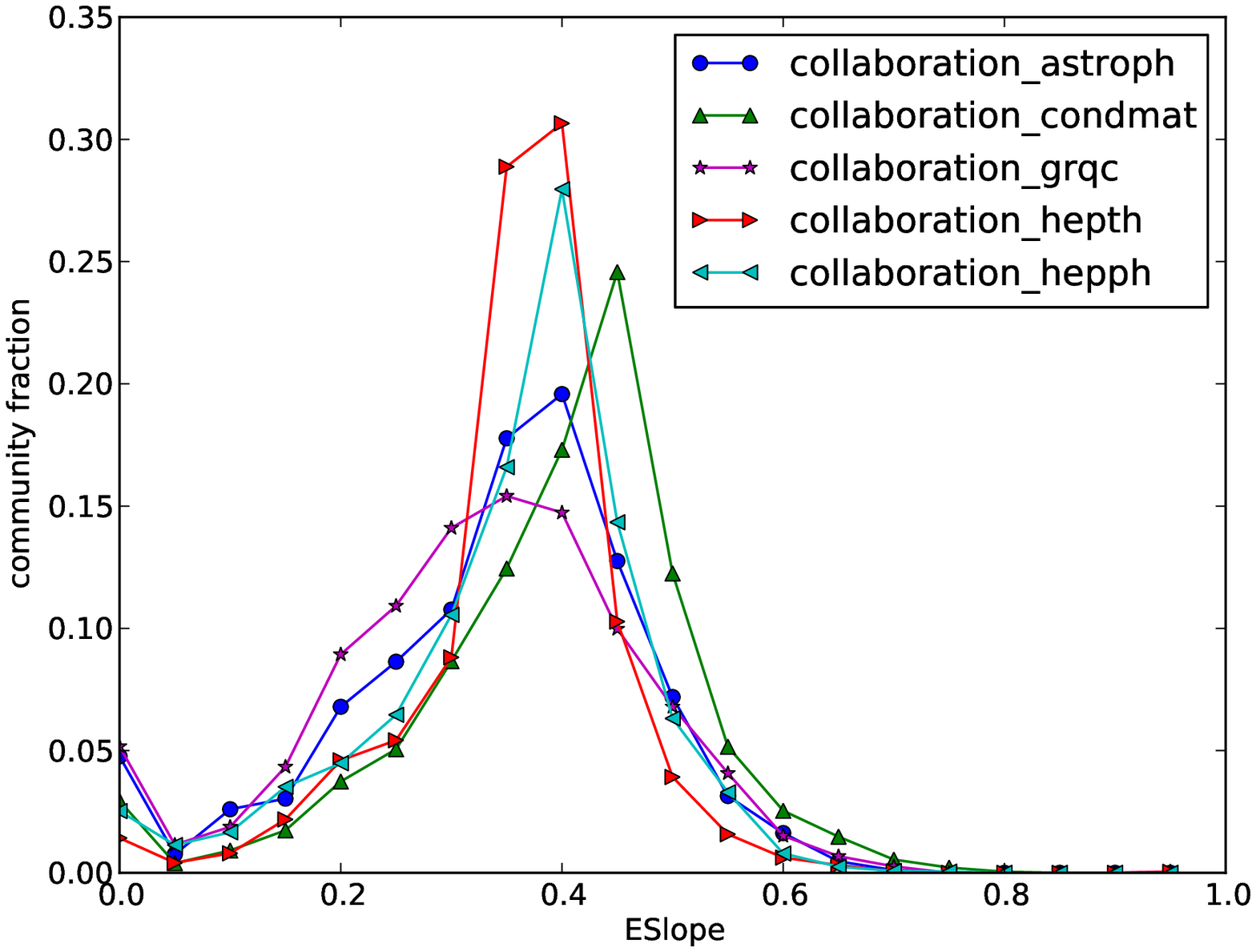}
}
\subfigure[ESlope\_email]
{\label{fig_eslope_email}
\includegraphics[width=2.5in]{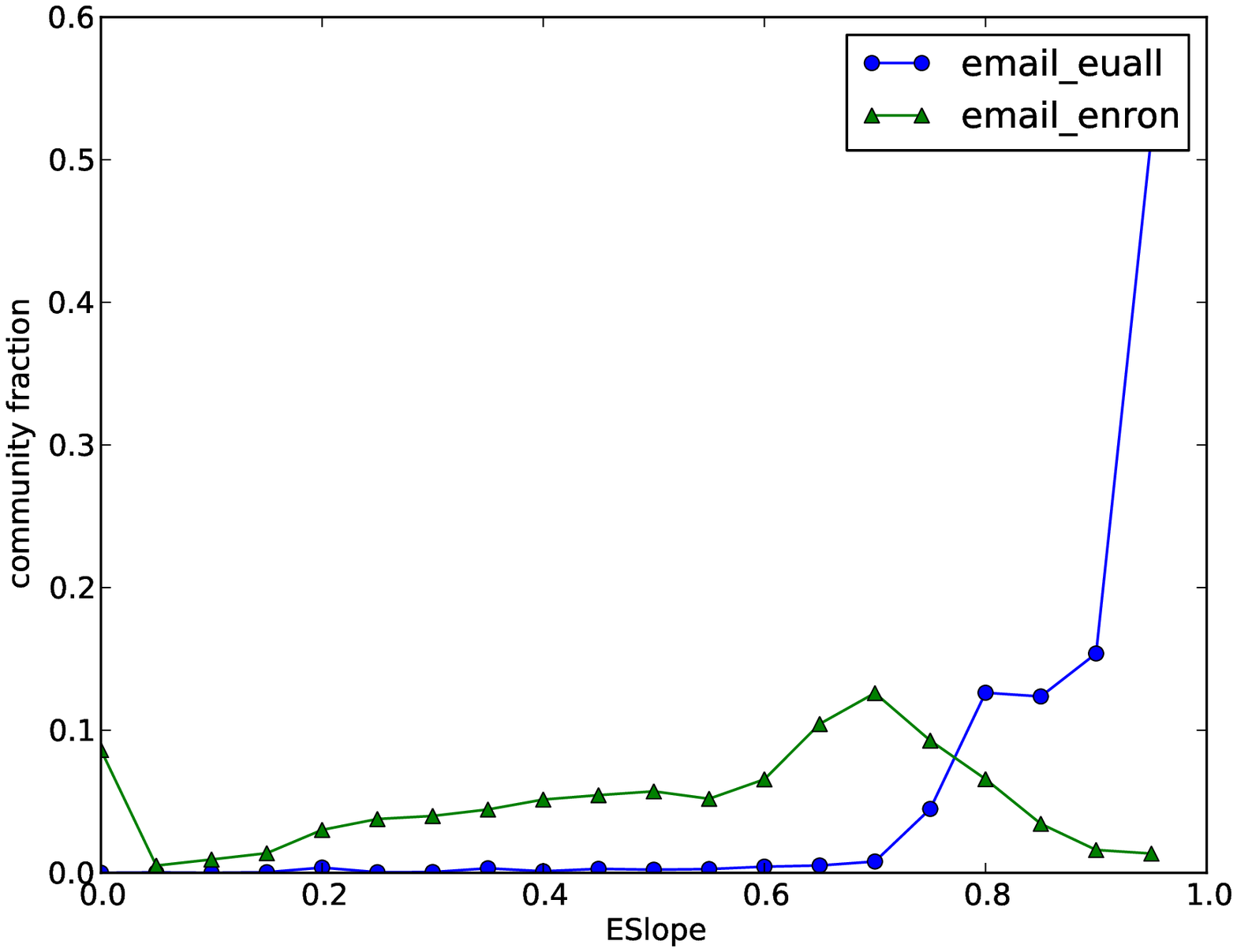}
}
\caption{Distribution of communities' ESlope}
\label{fig_eslope_distri}
\end{figure}

\begin{figure}
\centering
\subfigure[ISlope\_football]
{\label{fig_islope_football}
\includegraphics[width=2.5in]{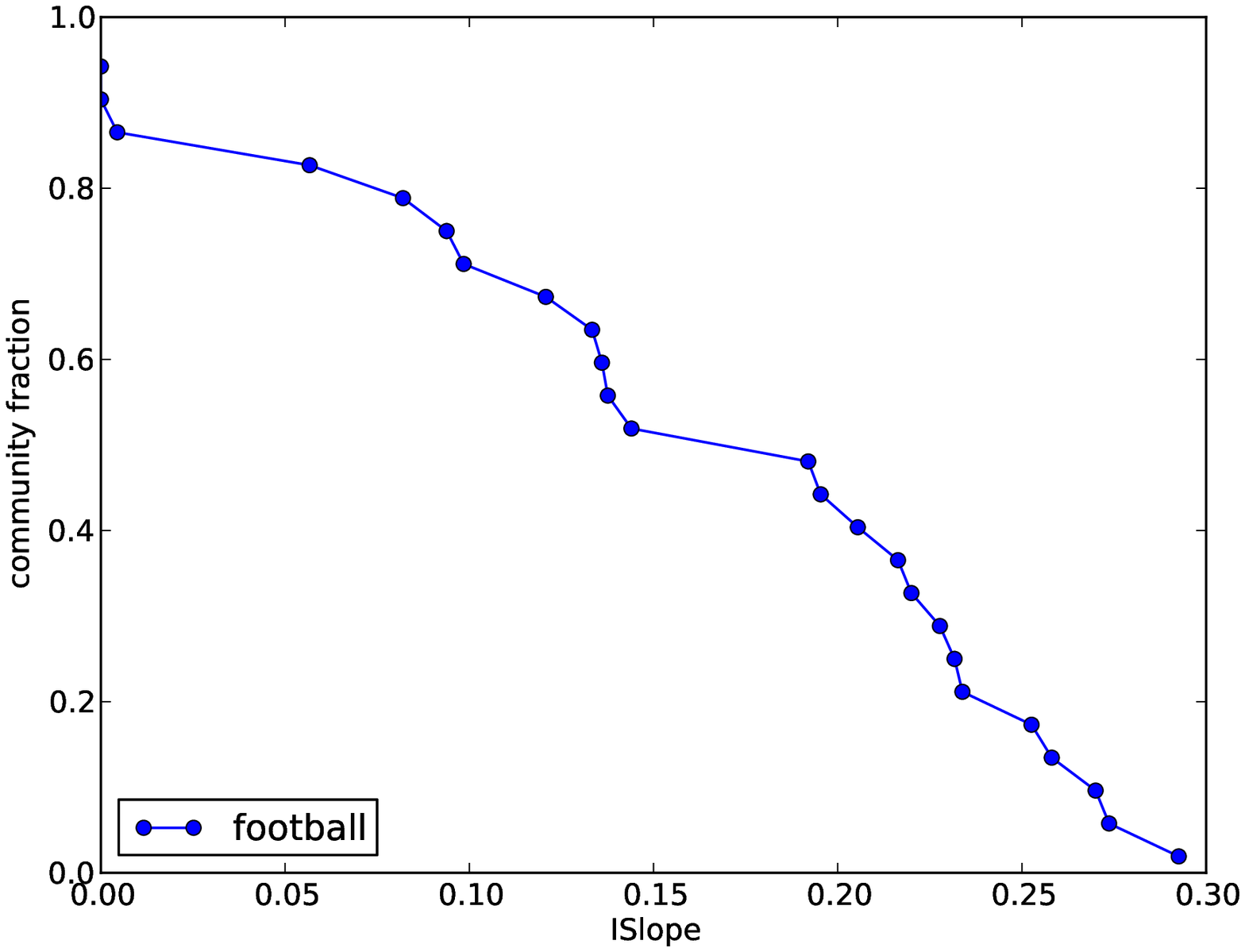}
}
\subfigure[ISlope\_citation]
{\label{fig_islope_citation}
\includegraphics[width=2.5in]{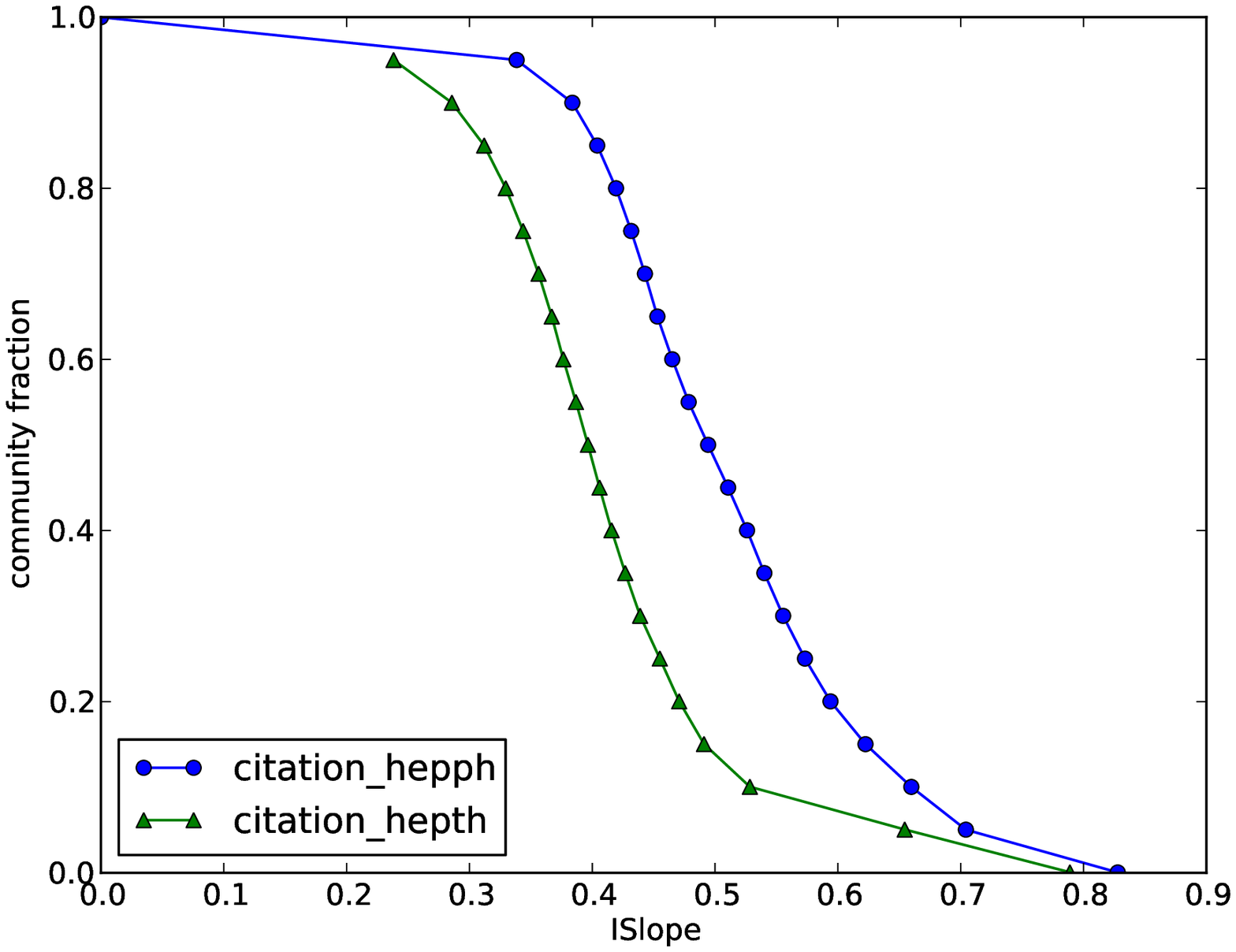}
}
\subfigure[ISlope\_collaboration]
{\label{fig_islope_collaboration}
\includegraphics[width=2.5in]{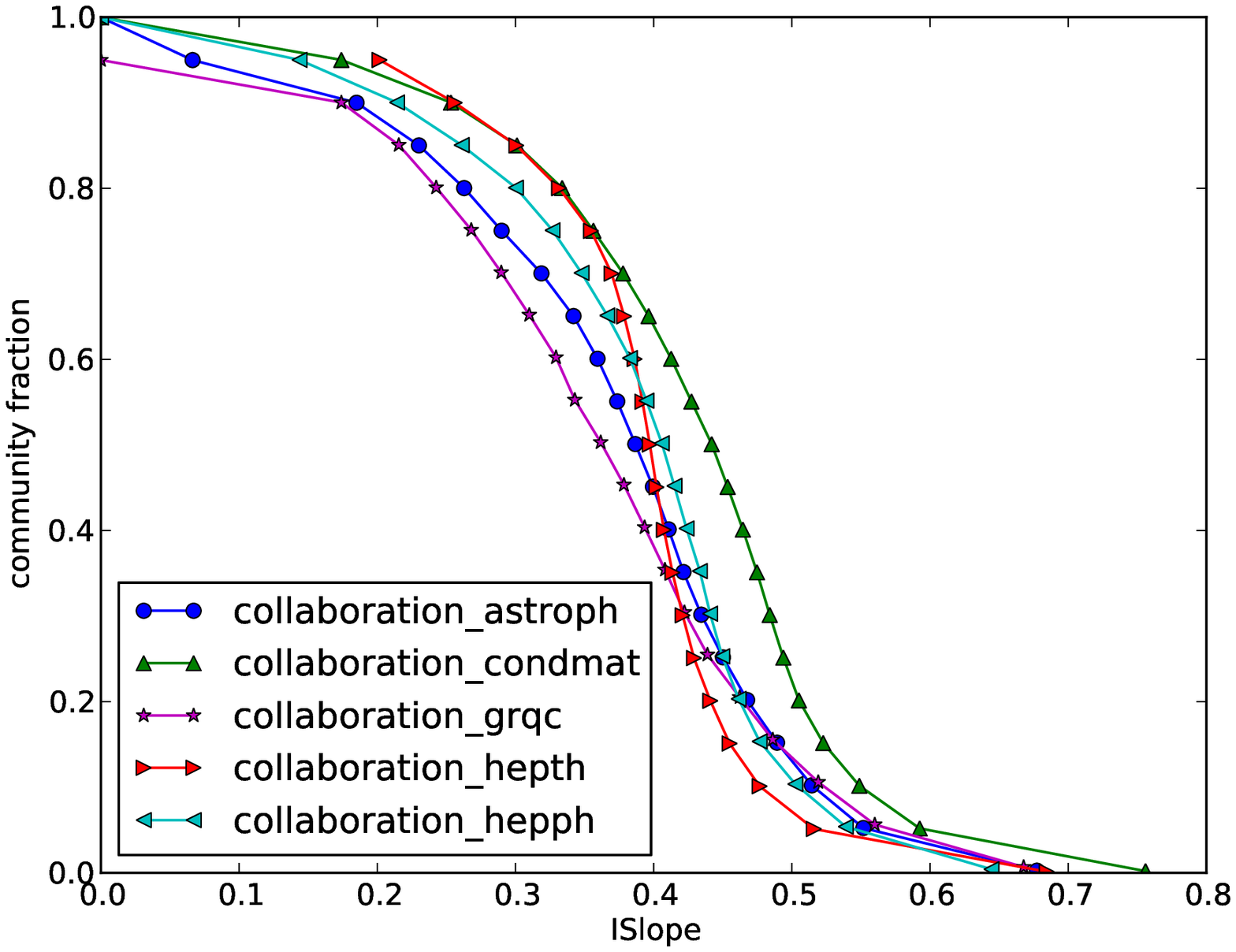}
}
\subfigure[ISlope\_email]
{\label{fig_islope_email}
\includegraphics[width=2.5in]{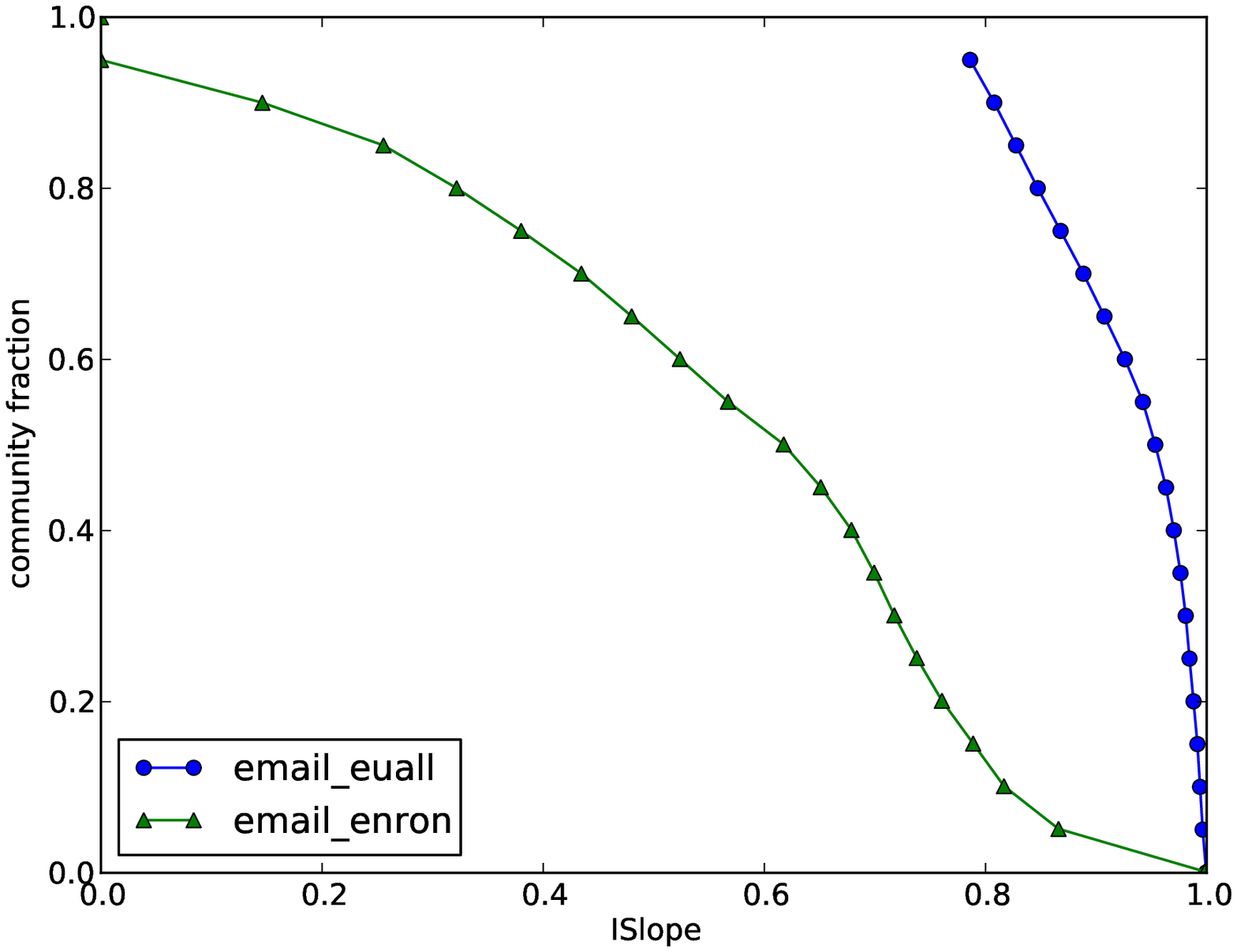}
}
\caption{Cumulative distribution of communities' ISlope}
\label{fig_islope_cumulative}
\end{figure}

\begin{figure}
\centering
\subfigure[ESlope\_football]
{\label{fig_eslope_football}
\includegraphics[width=2.5in]{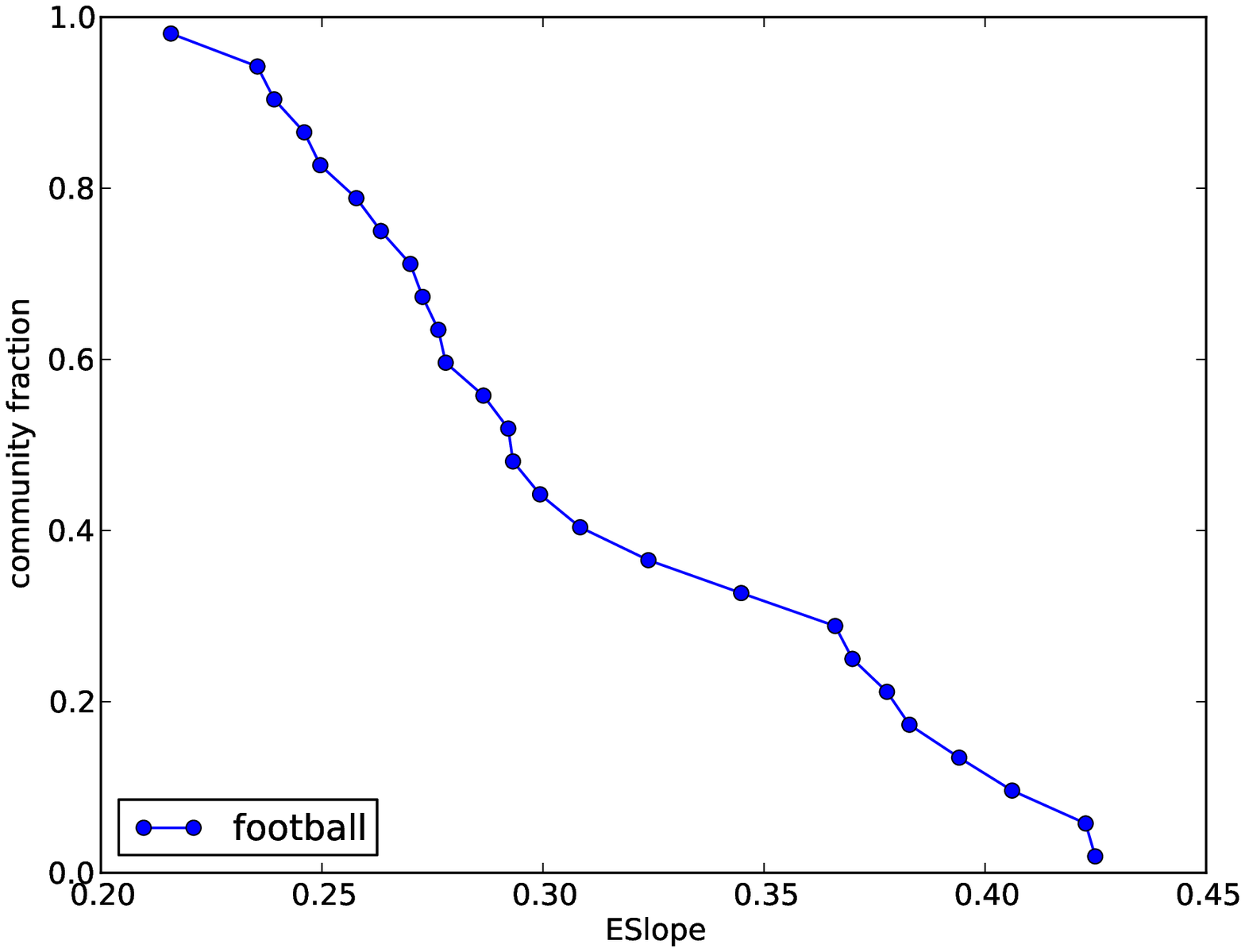}
}
\subfigure[ESlope\_citation]
{\label{fig_eslope_citation}
\includegraphics[width=2.5in]{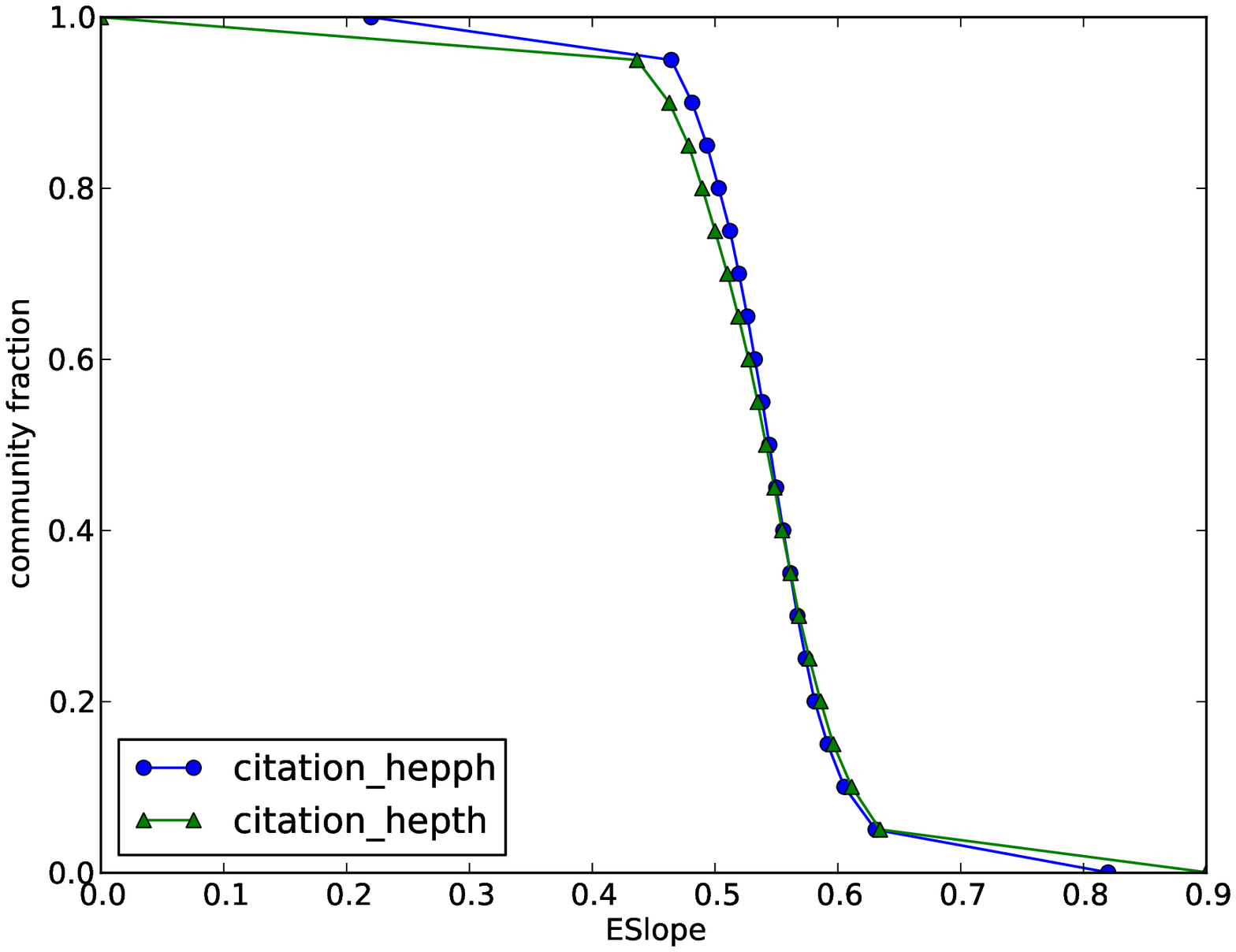}
}
\subfigure[ESlope\_collaboration]
{\label{fig_eslope_collaboration}
\includegraphics[width=2.5in]{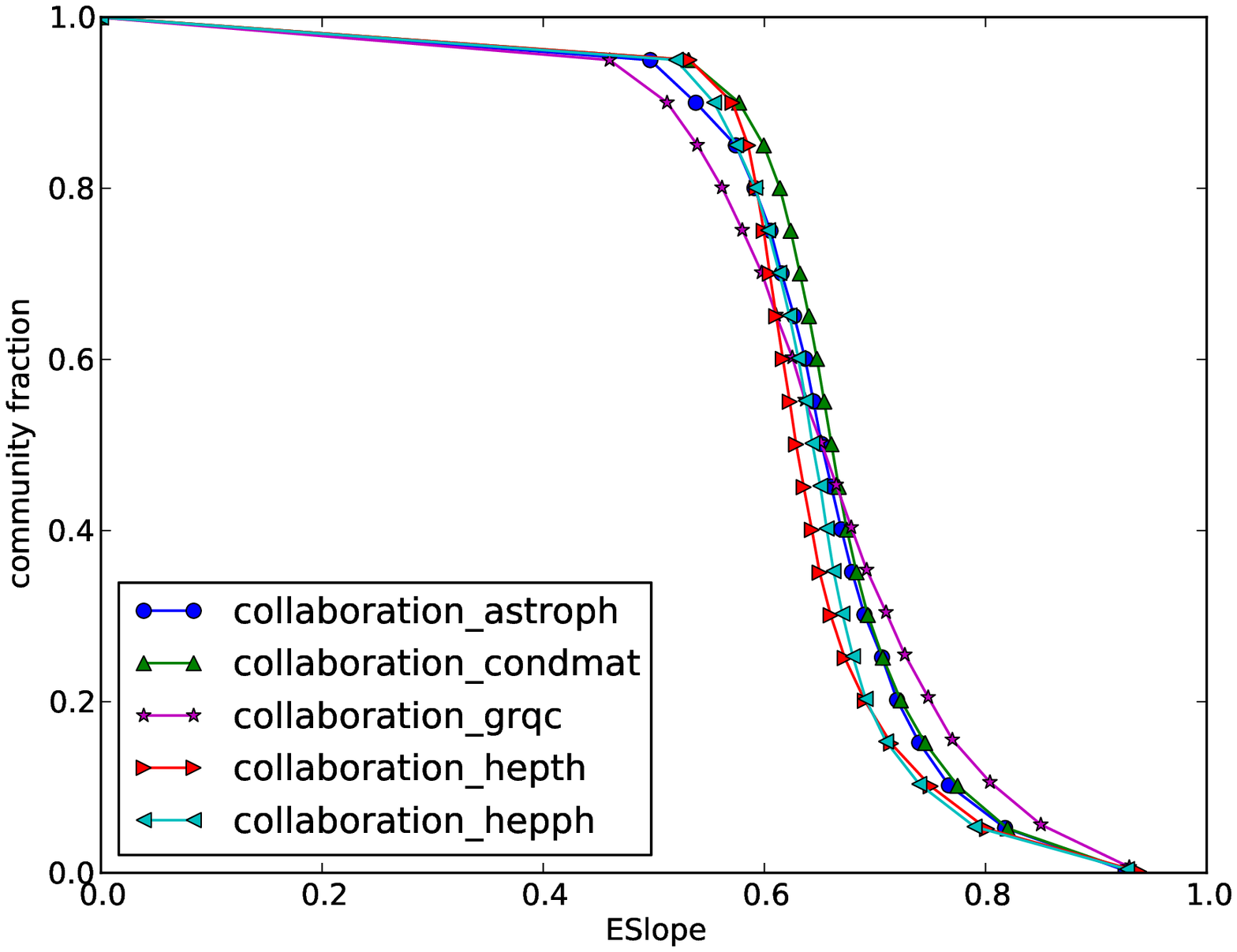}
}
\subfigure[ESlope\_email]
{\label{fig_eslope_email}
\includegraphics[width=2.5in]{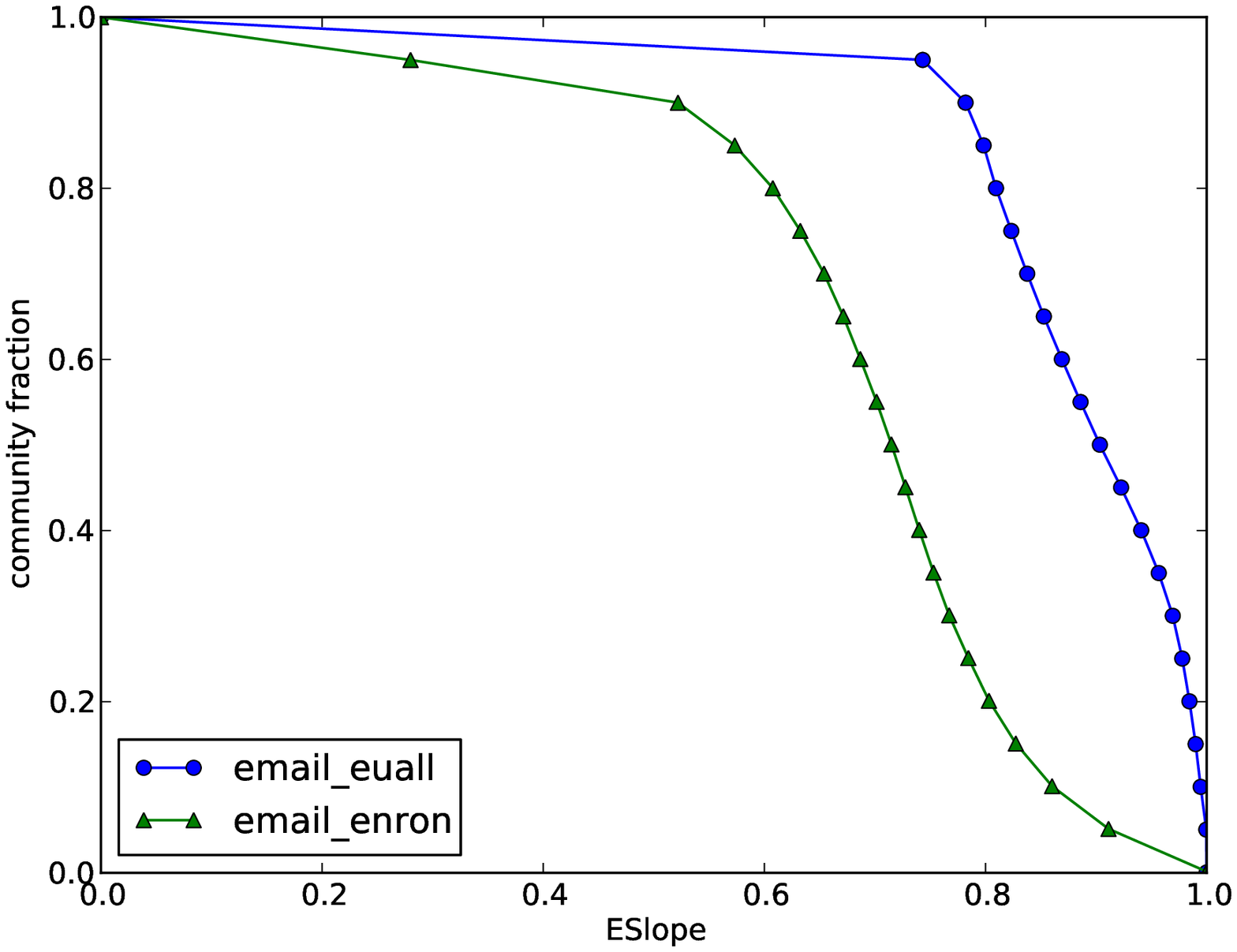}
}
\caption{Cumulative distribution of communities' ESlope}
\label{fig_eslope_cumulative}
\end{figure}

Besides the average values, we also report the distributions of the
ISlopes and ESlopes in figure \ref{fig_islope_distri} and figure
\ref{fig_eslope_distri} of all the communities of the real
networks. Figure \ref{fig_islope_cumulative} and figure
\ref{fig_eslope_cumulative} are the corresponding
cumulative distribution. 
By observing these figures, we know that:
\begin{itemize}
\item Most communities have a core/periphery structure, with a small
core in central positions and some low degree nodes in the
periphery.
\item The ISlopes largely determine the structure of the
communities.

\item There are indeed some typical thresholds at which
the distribution curve decreases sharply in most networks.
\item The typical values of ESlopes are more obvious than that of the
ISlopes in the citation and collaboration networks, in which, the
ESlopes of most communities lie in a very narrow interval.
\item Communities of the email-euall network have much larger
ISlopes and ESlopes in general.
\item The ISlopes and ESlopes of all the communities of the
citation and collaboration networks
 approximately follow a normal distribution.
\end{itemize}

\section{More General Properties}
\label{other_general_properties}

\begin{table}
\tbl{Statistics of communities. APL represents
average path length, D represents diameter, CCC represents community
clustering coefficient
and NCC represents the network clustering coefficient.
All the results except NCC are calculated by averaging the corresponding
property of all communities\label{table:general_community_data}
}
{
\begin{tabular} {|c|c|c|c|c|}
\hline
Network&APL&D&CCC&NCC\\
\hline
football&1.8&3.2&0.6&0.41\\
\hline
cit\_hepth&2.9&7&0.36&0.12\\
\hline
cit\_hepph&2.7&6.7&0.29&0.15\\
\hline
col\_astroph&2.2&4.5&0.71&0.32\\
\hline
col\_condmat&2.7&5.4&0.53&0.26\\
\hline
col\_grqc&2.4&4.7&0.51&0.63\\
\hline
col\_hepth&3.3&7.2&0.39&0.28\\
\hline
col\_hepph&2.8&5.9&0.65&0.66\\
\hline
email\_enron&2.2&4.1&0.39&0.085\\
\hline
email\_euall&2.3&3.5&0.0019&0.0042\\
\hline
\end{tabular}
}
\end{table}

In the last section, we show that the internal slope (ISlope) of a
community basically determines the structure of the community. In
this section, we study more general properties of the communities.
In particular, we consider the average distances, average diameters
and average clustering coefficients of all the communities in each
of the real networks, for which the results are given in Table
\ref{table:general_community_data}.

The distance between two nodes is defined as the number of ``hops"
in the network one needs to move from one given node to another
\cite{newman2004coauthorship}. Usually people are interested in the
average distances of the whole network~\cite{milgram1967small,newman2001structure,newman2001bscientific,travers1969experimental},
showing that most real networks have very
short average distances. In this section, we consider the average
distance between two nodes within a community, which represents the
number of ``hops" one needs to move from one node to another only
through members of the same community.

From Table \ref{table:general_community_data}, we have that, the
communities of each network have a small average distance. In
particular, the average distance of all the communities of the
collaboration network hepth reaches $3.3$, which is the largest
value of the average distances of all the communities for all the
networks studied in this paper. Besides, we also give the average
diameter of communities. The average diameter of all the communities
for each of the networks is between $3.2$ and $7.2$. This experiment
suggests a conjecture that: there is a {\it three degree separation
property of (true) communities} for many real networks. The
conjecture calls for further investigation, which may provide useful
information for understanding both true communities and communities
found by various algorithms.

 Clustering coefficient (or transitivity) has been a well
studied property for networks 
\cite{newman2001ascientific,newman2001structure,watts1998collective}.
It refers to the
phenomenon that the existence of ties between nodes $A$ and $B$ and
between nodes $B$ and $C$ implies a tie between $A$ and $C$. Given a
graph $G$, the clustering coefficient of $G$ is defined by:
\begin{equation}
C = \frac{3 \times \textrm{number of triangles on the graph}}
{\textrm{number of connected triples of vertices}}
\end{equation}

From table \ref{table:general_community_data}, we observe that most
communities of the networks have very large clustering coefficients
except for that of the email\_euall network, and that most small
communities found by our algorithm have larger clustering
coefficients than that of the corresponding original graphs.

However, in the collaboration network grqc, the clustering
coefficient of the original graph is $0.63$, but many small
communities we found have smaller clustering coefficients. In fact,
communities with clustering coefficients less than $0.6$ take up
more than $74\%$ of the communities in this network. To explain this
phenomenon, we count the triangles in the original graph and its
communities respectively. In the original graph, there are
$1,350,014$ triangles in all. If we divide the communities into two
groups, so that the first group consists of the ones having
clustering coefficients larger than $0.6$, and the second group
consists of the rest of the communities, then we discover that
communities in the first group have $3,306$ triangles on the
average, while communities in the second group contain $60$
triangles on the average. If we divide communities by clustering
coefficient $0.8$ as above, then the average numbers of triangles
appear in the communities in the first and the second classes are $
5,027$ and $147$ respectively. Therefore the triangles are unevenly
distributed in communities with a small number of communities
containing most of triangles of the network. The high clustering
coefficients are mainly caused by the small group of communities
which contain much larger number of triangles.

From table \ref{table:general_community_data}, we observe that
clustering coefficients of communities vary among different types of
networks. Communities in collaboration networks have higher
clustering coefficients than that of citation and email networks. In
the collaboration networks, two authors having common collaborators
are more likely to collaborate with each other in the future. In the
citation networks, an author citing a paper, tends to cite the
references of the paper, especially  when the references are from
the same topic. This explains the reason why collaboration networks
and citation networks have higher clustering coefficients.

Email networks have different patterns. Communities in email\_enron
network have average clustering coefficient $0.39$, at the same
time, the origin graph has clustering coefficient only $0.085$. In
this case, the communities found by our algorithm largely amplify
the clustering coefficients of the network. This means that although
the network has a small clustering coefficient, there are also
significantly many local structures of the network showing strong
cohesion among themselves. However communities in email\_euall
network has the lowest clustering coefficient (only $0.0019$). Both
its origin and communities have very small clustering coefficients.
In this case, most communities in this network are very similar to
star-like graphs which have clustering coefficients near $0$. This
local structure of the network is very much different from other
networks.

\section{Conclusions}
\label{conclusion} In this paper, we propose a methodology to
characterize and analyze the local structures and information of
real networks, which includes new notions of internal dominating
set, external dominating set, internal slope and external slope of a
community, and analysis of the distributions of internal and
external slopes, average distances, diameters, and clustering
coefficients of all communities for each of the real networks.

We implement experiments of our method on five collaboration
networks, two citation networks, two email networks and one
benchmark network.

The experiments show that: 1) The notions of internal dominating
ratio, external dominating ratio, internal slope and external slope
and clustering coefficients are essential characteristics to
understand the patterns and information of the communities of a
real network. 2) Different networks have different local structures
(or patterns). 3) Most communities of a real network have a small
internal dominating set and a small external dominating set,
although the communities may still very large. 4) The small
dominating set of a community keeps much of the information of the
community and more importantly the information of a community can
be extracted from the internal dominating set of the community. 5)
Both internal and external slopes of all the communities of a
network approximately follow a normal distribution for most real
networks. This means that typical communities of the networks have
both ISlopes and ESlopes in some small intervals, so that the
communities have similar patterns. 6) The internal slope (ISlope) of
a community basically determines the structure of the community. 7)
The result that communities have average distances less than or
equal to $ 3.3$, implies a general conjecture that there is a  $3$
degree separation phenomenon of true communities of most real
networks. 8) Normally, communities amplify the clustering
coefficients of the corresponding network. 9) If a reasonably good
algorithm fails to find communities that amplify clustering
coefficients of the network, then the communities explore special
structures of the network.

The discoveries above are significant in both understanding the
structures of networks, and in practical applications. Most
communities in real networks are not regular or star-like graphs,
but they usually appear with some central nodes with periphery
around forming a core/periphery structure. Such structure favors the
evolution of communities. A small set of nodes lead to the formation
and evolution of the communities. Our results also indicate that in
real communities,  a single node could rarely take absolute central
position as in star-like graphs, due to the reason that such
structures are highly unstable. Our analysis provides some intuitive
pictures of the rich communities of a network.

In best of our knowledge, this is the first time we can rigorously
analyze the characteristics and patterns, and extract information of
the communities of a real network, although there are already a huge
number of community detection algorithms in the literature.  The
significance of the research are three folds: 1) To understand the
local structures and connecting patterns of a network. 2) To extract
useful information from the communities of a network. 3) To help to
judge the community finding algorithms.

Our future project (in progress) is to understand the roles of the
small internal and external dominating sets in the formation and
evolution of communities, and to understand the mechanisms of the
patterns of the communities.

\bibliographystyle{acmsmall}
\bibliography{mybib1}

\begin{thebibliography}{}

\bibitem[\protect\citeauthoryear{Allan and Laskar}{Allan and
  Laskar}{1978}]{allan1978domination}
{\sc Allan, R.} {\sc and} {\sc Laskar, R.} {1978}.
\newblock {On domination and independent domination numbers of a graph}.
\newblock {\em {Discrete Mathematics }\/}~{\em {23},\/}~{2}, {73--76}.

\bibitem[\protect\citeauthoryear{Andersen, Chung, and Lang}{Andersen
  et~al\mbox{.}}{2006}]{andersen2006local}
{\sc Andersen, R.}, {\sc Chung, F.}, {\sc and} {\sc Lang, K.} 2006.
\newblock Local graph partitioning using pagerank vectors.
\newblock In {\em 47th Annual IEEE Symposium on Foundations of Computer
  Science, 2006. FOCS'06.} IEEE, 475--486.

\bibitem[\protect\citeauthoryear{Barab{\'a}si and Albert}{Barab{\'a}si and
  Albert}{1999}]{barabasi1999emergence}
{\sc Barab{\'a}si, A.} {\sc and} {\sc Albert, R.} 1999.
\newblock Emergence of scaling in random networks.
\newblock {\em Science\/}~{\em 286,\/}~5439, 509.

\bibitem[\protect\citeauthoryear{Bavelas}{Bavelas}{1948}]{bavelas1948mathemati%
cal}
{\sc Bavelas, A.} 1948.
\newblock A mathematical model for group structures.
\newblock {\em Human Organization\/}~{\em 7,\/}~3, 16--30.

\bibitem[\protect\citeauthoryear{Clauset, Newman, and Moore}{Clauset
  et~al\mbox{.}}{2004}]{clauset2004finding}
{\sc Clauset, A.}, {\sc Newman, M.}, {\sc and} {\sc Moore, C.} 2004.
\newblock Finding community structure in very large networks.
\newblock {\em Physical Review E\/}~{\em 70,\/}~6, 066111.

\bibitem[\protect\citeauthoryear{Clauset, Shalizi, and Newman}{Clauset
  et~al\mbox{.}}{2009}]{clauset2009power}
{\sc Clauset, A.}, {\sc Shalizi, C.~R.}, {\sc and} {\sc Newman, M. E.~J.}
  {2009}.
\newblock {Power-Law Distributions in Empirical Data}.
\newblock {\em {SIAM Review}\/}~{\em {51},\/}~{4}, {661--703}.

\bibitem[\protect\citeauthoryear{De~Nooy, Mrvar, and Batagelj}{De~Nooy
  et~al\mbox{.}}{2011}]{de2011exploratory}
{\sc De~Nooy, W.}, {\sc Mrvar, A.}, {\sc and} {\sc Batagelj, V.} 2011.
\newblock {\em Exploratory Social Network Analysis With Pajek}. Vol.~34.
\newblock Cambridge Univ Pr.

\bibitem[\protect\citeauthoryear{Fortunato}{Fortunato}{2010}]{fortunato2010com%
munity}
{\sc Fortunato, S.} 2010.
\newblock Community detection in graphs.
\newblock {\em Physics Reports\/}~{\em 486,\/}~3-5, 75--174.

\bibitem[\protect\citeauthoryear{Freeman}{Freeman}{1979}]{freeman1979centralit%
y}
{\sc Freeman, L.} 1979.
\newblock Centrality in social networks conceptual clarification.
\newblock {\em Social Networks\/}~{\em 1,\/}~3, 215--239.

\bibitem[\protect\citeauthoryear{Girvan and Newman}{Girvan and
  Newman}{2002}]{girvan2002community}
{\sc Girvan, M.} {\sc and} {\sc Newman, M.} 2002.
\newblock Community structure in social and biological networks.
\newblock {\em Proceedings of the National Academy of Sciences\/}~{\em
  99,\/}~12, 7821.

\bibitem[\protect\citeauthoryear{Goyal, Van Der~Leij, and
  Moraga-Gonz{\'a}lez}{Goyal et~al\mbox{.}}{2006}]{goyal2006economics}
{\sc Goyal, S.}, {\sc Van Der~Leij, M.}, {\sc and} {\sc Moraga-Gonz{\'a}lez,
  J.} 2006.
\newblock Economics: An emerging small world.
\newblock {\em Journal of Political Economy\/}~{\em 114,\/}~2, 403--412.

\bibitem[\protect\citeauthoryear{Guimera and Amaral}{Guimera and
  Amaral}{2005}]{guimera2005functional}
{\sc Guimera, R.} {\sc and} {\sc Amaral, L.} 2005.
\newblock Functional cartography of complex metabolic networks.
\newblock {\em Nature\/}~{\em 433,\/}~7028, 895--900.

\bibitem[\protect\citeauthoryear{Haynes, Hedetniemi, and Slater}{Haynes
  et~al\mbox{.}}{1998}]{haynes1998funcamentals}
{\sc Haynes, T.}, {\sc Hedetniemi, S.}, {\sc and} {\sc Slater, P.} 1998.
\newblock {\em Fundamentals of domination in graphs}.

\bibitem[\protect\citeauthoryear{Leskovec, Kleinberg, and Faloutsos}{Leskovec
  et~al\mbox{.}}{2007}]{leskovec2007graph}
{\sc Leskovec, J.}, {\sc Kleinberg, J.}, {\sc and} {\sc Faloutsos, C.} 2007.
\newblock Graph evolution: Densification and shrinking diameters.
\newblock {\em ACM Transactions on Knowledge Discovery from Data (TKDD)\/}~{\em
  1,\/}~1, 2.

\bibitem[\protect\citeauthoryear{Leskovec, Lang, Dasgupta, and
  Mahoney}{Leskovec et~al\mbox{.}}{2009}]{leskovec2009community}
{\sc Leskovec, J.}, {\sc Lang, K.}, {\sc Dasgupta, A.}, {\sc and} {\sc Mahoney,
  M.} 2009.
\newblock Community structure in large networks: Natural cluster sizes and the
  absence of large well-defined clusters.
\newblock {\em Internet Mathematics\/}~{\em 6,\/}~1, 29--123.

\bibitem[\protect\citeauthoryear{Li and Peng}{Li and
  Peng}{2011}]{li2011community}
{\sc Li, A.} {\sc and} {\sc Peng, P.} 2011.
\newblock Community structures in classical network models.
\newblock {\em Internet Mathematics\/}~{\em 7,\/}~2, 81--106.

\bibitem[\protect\citeauthoryear{Li and Peng}{Li and Peng}{2012}]{li2012smalla}
{\sc Li, A.} {\sc and} {\sc Peng, P.} 2012.
\newblock The small-community phenomenon in networks.
\newblock {\em Math. Struct. in Comp. Science\/}~{\em 22}, 1--35.

\bibitem[\protect\citeauthoryear{Milgram}{Milgram}{1967}]{milgram1967small}
{\sc Milgram, S.} 1967.
\newblock The small world problem.
\newblock {\em Psychology Today\/}~{\em 2,\/}~1, 60--67.

\bibitem[\protect\citeauthoryear{Nacher and Akutsu}{Nacher and
  Akutsu}{2012}]{jose2012dominating}
{\sc Nacher, J.~C.} {\sc and} {\sc Akutsu, T.} 2012.
\newblock Dominating scale-free networks with variable scaling exponent:
  heterogeneous networks are not difficult to control.
\newblock {\em New Journal of Physics\/}~{\em 14,\/}~7, 073005.

\bibitem[\protect\citeauthoryear{Newman}{Newman}{2001a}]{newman2001ascientific}
{\sc Newman, M.} 2001a.
\newblock Scientific collaboration networks. i. network construction and
  fundamental results.
\newblock {\em Physical Review E\/}~{\em 64,\/}~1, 016131.

\bibitem[\protect\citeauthoryear{Newman}{Newman}{2001b}]{newman2001structure}
{\sc Newman, M.} 2001b.
\newblock The structure of scientific collaboration networks.
\newblock {\em Proceedings of the National Academy of Sciences\/}~{\em
  98,\/}~2, 404.

\bibitem[\protect\citeauthoryear{Newman}{Newman}{2004a}]{newman2004coauthorshi%
p}
{\sc Newman, M.} 2004a.
\newblock Coauthorship networks and patterns of scientific collaboration.
\newblock {\em Proceedings of the National Academy of Sciences of the United
  States of America\/}~{\em 101,\/}~Suppl 1, 5200.

\bibitem[\protect\citeauthoryear{Newman}{Newman}{2004b}]{newman2004detecting}
{\sc Newman, M.} 2004b.
\newblock Detecting community structure in networks.
\newblock {\em The European Physical Journal B-Condensed Matter and Complex
  Systems\/}~{\em 38,\/}~2, 321--330.

\bibitem[\protect\citeauthoryear{Newman et~al\mbox{.}}{Newman
  et~al\mbox{.}}{2001}]{newman2001bscientific}
{\sc Newman, M.} {\sc et~al\mbox{.}} 2001.
\newblock Scientific collaboration networks. ii. shortest paths, weighted
  networks, and centrality.
\newblock {\em Physical Review E\/}~{\em 64,\/}~1; PART 2, 16132--16132.

\bibitem[\protect\citeauthoryear{Nicosia, Criado, Romance, Russo, and
  Latora}{Nicosia et~al\mbox{.}}{2012}]{nicosia2012controlling}
{\sc Nicosia, V.}, {\sc Criado, R.}, {\sc Romance, M.}, {\sc Russo, G.}, {\sc
  and} {\sc Latora, V.} 2012.
\newblock Controlling centrality in complex networks.
\newblock {\em Scientific Reports\/}~{\em 2}.

\bibitem[\protect\citeauthoryear{Rogers}{Rogers}{1974}]{rogers1974sociometric}
{\sc Rogers, D.} 1974.
\newblock Sociometric analysis of interorganizational relations: Application of
  theory and measurement.
\newblock {\em Rural Sociology\/}.

\bibitem[\protect\citeauthoryear{Travers and Milgram}{Travers and
  Milgram}{1969}]{travers1969experimental}
{\sc Travers, J.} {\sc and} {\sc Milgram, S.} 1969.
\newblock An experimental study of the small world problem.
\newblock {\em Sociometry\/}, 425--443.

\bibitem[\protect\citeauthoryear{Watts and Strogatz}{Watts and
  Strogatz}{1998}]{watts1998collective}
{\sc Watts, D.} {\sc and} {\sc Strogatz, S.} 1998.
\newblock Collective dynamics of `small-world' networks.
\newblock {\em Nature\/}~{\em 393,\/}~6684, 440--442.

\end{thebibliography}
\end{document}